\documentclass[%
 aip,
 jmp,%
 amsmath,amssymb,
 reprint,%
]{revtex4-1}

\usepackage{color}
\usepackage{graphicx}
\usepackage{dcolumn}
\usepackage{bm}
\usepackage{cancel}
\usepackage[normalem]{ulem}

\newcommand{\nobracket}{}
\newcommand{\nocomma}{}
\newcommand{\tmmathbf}[1]{\ensuremath{\boldsymbol{#1}}}
\newcommand{\tmop}[1]{\ensuremath{\operatorname{#1}}}

\newcommand{\red}{black}
\newcommand{\sns}[1]{}
\newcommand{\snseq}[1]{}

\begin{document}

\preprint{AIP/123-QED}

\title[A hierarchy of generalized kinetic equations]{A consistent hierarchy of generalized kinetic equation approximations to the master equation applied to surface catalysis}

\author{Gregory J. Herschlag}
 \altaffiliation[]{Mathematics Department, Duke University, Durham, NC, 27708, USA}
 \email{gjh@math.duke.edu}
\author{Sorin Mitran}%
 \altaffiliation[]{Department of Mathematics, University of North Carolina at Chapel Hill,  Chapel Hill, NC, 27599, USA}
\author{Guang Lin}%
 \altaffiliation[]{Department of Mathematics, Purdue University,  West Lafayette, IN 47907, USA\\
  School of Mechanical Engineering, Purdue University, West Lafayette, IN 47907, USA}

\date{\today}

\begin{abstract}
We develop a hierarchy of approximations to the master equation for systems that exhibit translational invariance and finite-range spatial correlation. Each approximation within the hierarchy is a set of ordinary differential equations that {\color{\red} considers spatial correlations of varying lattice distance; the assumption is that the full system will have finite spatial correlations and thus the behavior of the models within the hierarchy will approach that of the full system. \sns{partially models the full system spatial correlation range. In principle, the hierarchy converges to the master equation, as evidenced by} We provide evidence of this convergence in the context of } 
one- and two-dimensional numerical examples. Lower levels within the hierarchy {\color{\red} that consider shorter spatial correlations,} are
shown to be up to three orders of magnitude faster than traditional kinetic Monte Carlo
methods (KMC) for one-dimensional systems, while predicting similar system dynamics
and steady states as KMC methods.
We {\color{\red} then test the hierarchy on a \sns{this methodology to model}} two-dimensional {\color{\red} model for the} oxidation of CO on RuO2(110),
showing that low-order truncations of the hierarchy efficiently capture the essential system dynamics.
{\color{\red} By considering sequences of models in the hierarchy that account for longer spatial correlations, \sns{Results from successive truncations of the hierarchy} successive model predictions }may be used to establish
empirical approximation {\color{\red} of} error estimates.
The hierarchy may be thought of as a class of generalized phenomenological kinetic models since 
each element of the hierarchy approximates the master equation and the lowest level in the hierarchy 
is identical to {\color{\red} a} simple existing phenomenological kinetic models.
\end{abstract}

\pacs{82.40.-g, 82.65.+r, 82.20.-w, 82.40.-g, 82.30.Vy}
\maketitle

\section{Introduction}

Computational modeling plays an increasingly important role in the characterization and understanding of a broad range of elementary chemical transformations relevant to catalytic processes.
Such catalytic chemical reactions can be described by microscopic kinetic models such as the coarse-grained lattice kinetic Monte Carlo (KMC) method \cite{Jansen}, a computational simulation of the time evolution of some stochastic process.
Typically, simulation of processes arising in catalytic chemistry are carried out based upon rates for adsorption, reaction, desorption, and diffusion that are obtained from experiments, density functional theory (DFT), and transition state theory (TST).
If a system exhibits significant transport, hybrid methods \cite{MeiLin:11, BalterLinTartakovsky:11} for heterogeneous reaction kinetics can be constructed, which combine KMC for the chemical kinetics with finite difference methods for the continuum-level heat and mass transfer.
Underlying all models and algorithms for determining reaction dynamics is the master equation.
The master equation describes the evolution of a multivariate probability distribution function (PDF) for finding a surface in any given state \cite{Gillespie:92, Karlin, Jansen}.  The master equation, however is an infinite dimensional ordinary differential equation which cannot be solved exactly, and thus a number of techniques have been developed for finding approximate solutions.

One class of computational methods for catalytic processes hypothesize \emph{ad hoc} rate equations, derived by physical reasoning, to construct phenomenological kinetic\cite{Kotomin:96, Froment:05} (PK) models of surface processes.
Such PK models start from an idealized surface geometry for binding sites and site connections. 
For example, on a (110) idealized surface, {\color{\red}\sns{there}} one can define bridge and cus sites connected by a square lattice, as shown in Figure \ref{fig:surfaces}.
The models track the probability of finding a site of given type bound to a particular molecule, and use a maximum-entropy/well-mixed assumption to reconstruct spatially correlated information.
The well-mixed assumption on surfaces can often fail, and there are many examples in which a given kinetic model fits one set of data well, but fails with additional test data\cite{Temeletal:07}.

There have been two methodologies that have been developed to alleviate the deficiency of the PK models: generalized phenomenological kinetic (gPK) models and kinetic Monte Carlo (KMC) simulation.   GPK models are similar to PK models however they add evolution equations for spatial correlations.  For surface catalysis, the gPK framework was introduced in the 1990's by Mai, Kuzovkov, and von Niessen and others \cite{Mai:93,Mai:93b,Mai:93c}, and since then this work has both been extended \cite{Mai:94,Mai:94b,Mai:96,Mai:97,Kuzovkov:98,Dickman:99,Cortes:01} and applied \cite{DeDecker:02,Baker:10,Baker:11,Jansen,Johnston:12,Baker:13,Simpson:13} in a variety of ways.  The primary idea behind these works is to present nested sets of kinetic equations for the probability of finding a collection of $k$ surface sites in a particular configuration.  The dynamics for the $k$ collections can only be determined exactly if the state of the $k+1$ collections are known, which is analogous to the BBGKY hierarchy of statistical physics.  In principle this leads to an infinite system of equations.  In practice the nested chain is approximated through a truncation closure at some level in the hierarchy.

Despite the advances in gPK modeling, kinetic Monte Carlo (KMC) simulations are now often the method of choice that is used to determine and verify the surface dynamics and steady state values in surface catalysis problems \cite{Hansen:99, Zeigarnik:03, Kieken:04, Mei:06,ReuterScheffler:06,Rogal:07, Herdrich:12, Liu:13,Hoffmann:13, Callejas:13, Gibson:14}.  KMC algorithms are stochastic realizations of the surface dynamics that probabilistically update the state of some surface with large but finite size.
The advantage of KMC algorithms { when compared to PK models} is demonstrated in ref. \onlinecite{Temeletal:07}, in which the authors demonstrate the break down of PK models when compared to KMC predictions.
KMC simulations are, however, far more expensive than both PK and {\color{\red} low level} gPK models, and rely on slow statistical averaging for predicting desired observable quantities.

Because the computational cost associated with KMC simulations is significantly larger than low level hierarchy members of gPK models, it is worth investigating why the KMC simulations have become the method of choice when determining solutions to the master equation dynamics.  
 Possible motivations include: (1) {\color{\red} the simplicity of KMC algorithms \sns{simpler coding of KMC methods}}, (2) advances in computational power that make KMC simulations feasible, and (3) {\color{\red} straight forward convergence testing of KMC predictions\sns{ simpler convergence testing in KMC}}. Elaborating on point (3), KMC simulations allow for convergence tests by taking larger surface domains {\color{\red} and more statistical samples} while using the same implementation, {\color{\red}where\sns{ }as} gPK convergence tests require inclusion of {\color{\red} new\sns{higher}} elements within the hierarchy, with the associated additional coding effort. We know of no work (see, for example refs. \cite{Mai:93,Mai:93b,Mai:93c,Mai:94,Mai:94b,Mai:96,Mai:97,Kuzovkov:98,Dickman:99,Cortes:01}) in which the nested hierarchy is built beyond time evolution of pair correlations{\color{\red}\sns{, with}; in these works} triplet 
correlations {\color{\red} are} approximated from pairwise information. The problem of formulating a general
procedure to construct higher-order hierarchy truncations is still unsolved.  This means that although gPK techniques may be used to approximate the master equation, there is no methodology to test if this approximation has converged to the correct dynamics other than comparing the results with KMC simulation.  If gPK models are to become viable techniques in determining accurate approximations to the master equation, there must come along with them a generalized methodology for constructing arbitrary elements of the hierarchy so that convergence may be examined.  As shown in the supplemental material\cite{supmat}, it is possible to obtain an inconsistent model through an inappropriate triplet closure, further highlighting the need for a formulation of gPK models that incorporates a method for testing convergence of successive truncations of the approximation hierarchy.

To accomplish this formulation, the present work will take a slightly different approach to the nested schemes. Similarly to the nested gPK schemes we will also arrive at a generalized hierarchy of phenomenological kinetic models.  Our methodology begins with the observation that the simple PK models are probability distributions of a $(1\times1)$ group of sites on the surface.  We will denote an arbitrary grouping of sites \mbox{$(m\times n)$} sites as a `tile' on the surface and seek to determine the kinetics of the probability distribution of surface states on a tile.
We note that the $(2\times1)$ tiling is identical to the nested scheme presented in ref. \onlinecite{Mai:93} (for surfaces made up of a single site type), and reiterate that the triplet nested scheme{\color{\red}, mentioned but not studied in ref. \onlinecite{Mai:93c},} is different than the $(3\times 1)$ tiling scheme (see the supplemental material \cite{supmat}).
The novel aspect of the present work will be the generalized construction of members of the hierarchy which will provide consistent and increasingly accurate dynamics for larger tilings.  Such a framework has the potential to allow the gPK models {\color{\red}\sns{to}} not {\color{\red}to} rely on KMC simulation to test for accuracy, but rather to remain self contained by comparing the results from smaller tile dynamics to larger tile dynamics, with the hope that the scheme will converge with significant computational savings.  

We begin this work by exposing the formalism of the tiling idea in Section 2, and provide two examples of how to construct a set of ODEs within the tiling framework.  The second of these examples considers the oxidation of CO on a face centered cubic structure's (110) surface (also considered in refs. \onlinecite{ReuterScheffler:06, Temeletal:07}).
In Section 3 we provide numerical evidence that supports and verifies the formalism based on a 1D and 2D uniform surface with a square lattice.
In Section 4 we test the kinetic models resulting from site ($1\times1$) and pair ($2\times1$) tilings for the surface catalysis problem of CO oxidation demonstrated in refs. \onlinecite{ReuterScheffler:06, Temeletal:07}{\color{\red} \sns{;}. Although previous gPK studies have mentioned the ability to consider lattices with different site types, we know of no existing work that has formally studied the case of regular lattices with distinct site types.  We note, however, that there have been studies on lattices with randomly active/inactive site \cite{Mai:93b} as well as disordered heterogenous lattices \cite{Cortes:01}.\sns{w}W}e demonstrate that the pair tiling significantly reduces errors made in the PK formalism of ref. \onlinecite{Temeletal:07}.  We next examine the idea of having a mixed tiling scheme and show that improved accuracy may be efficiently obtained  within this mixed tiling hierarchy.  We present results for this mixed tiling and demonstrate that it better captures the dynamics predicted by KMC simulation, providing the possibility for a search algorithm in the tiling hierarchy that remains computationally inexpensive.

\section{The master equation and approximation}
\subsection{Problem formulation}
\begin{figure}
\mbox{\includegraphics[trim=10mm 10mm 20mm 10mm, height=5cm]{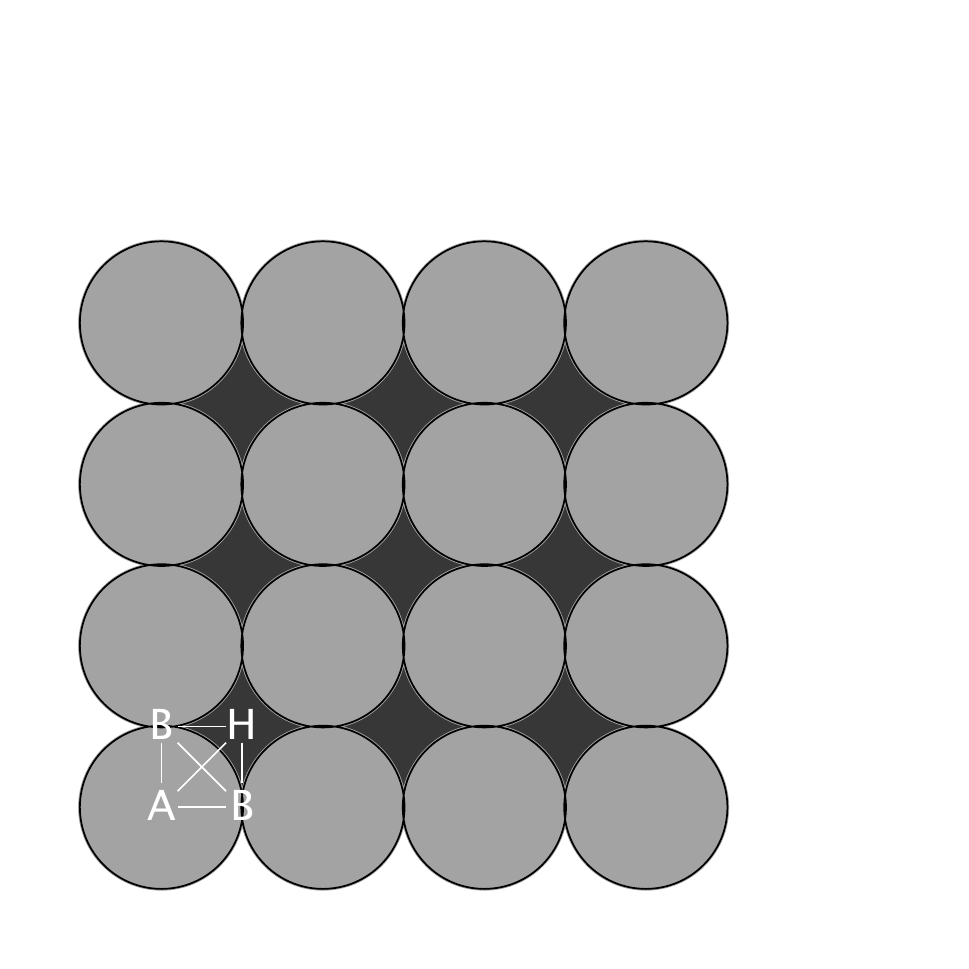}(a)}
\mbox{\includegraphics[trim=10mm 10mm 20mm 10mm, height=5cm]{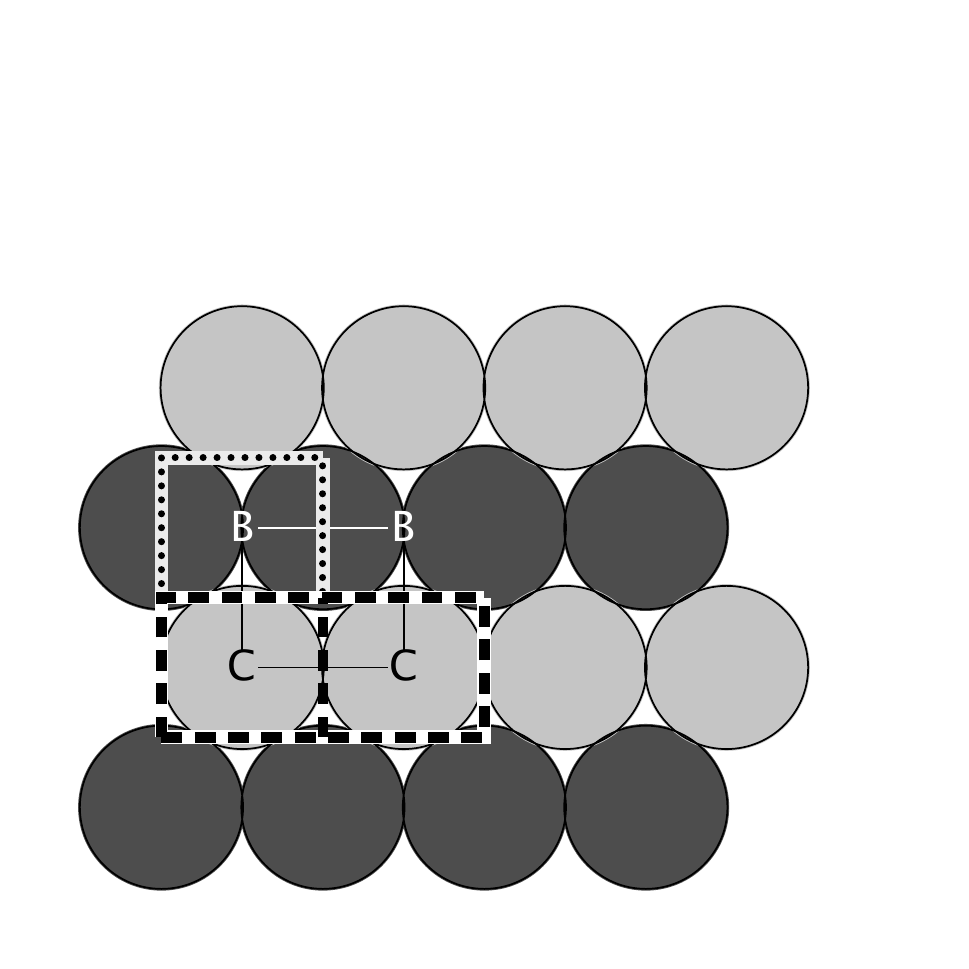}(b)}
\caption{The figure labeled (a) shows an idealized (100) surface with lattice sites: atop (A), bridge (B) and hollow (H).  The figure labeled (b) shows an idealized (110) surface with lattice sites: bridge (B) and cus (C); the two colorations represent different surface heights of the molecules comprising the layer.  When oxygen adsorbs on to the surface, it will dissociate and split into two adjacent sites.  Connections defining these adjacent sites are shown via the lines connecting the sites.  In (b) we display two tiles from distinct tiling schemes; the box surrounding the bridge site reduces the kinetic model to a traditional phenomenological kinetic model, whereas the box surrounding the two cus sites is a tile taken from a $(2\times1)$ tiling scheme.
}
\label{fig:surfaces}
\end{figure}

Consider a surface $\Omega$ made of $N\times N$ sites, each labeled as particular type from a set $\mathcal{T}$.
For example on an idealized (100) surface there are atop, bridge, and 4-fold hollow sites and so $\mathcal{T}=\{$atop, bridge, hollow$\}$, whereas on a (110) crystal surface there are bridge and cus site types  and so $\mathcal{T}=\{$bridge, cus$\}$ (see Figure \ref{fig:surfaces}).
Each site on the surface may be in a particular state and we will call the set of possible states $\mathcal{S}$.
For example, in the surface oxidation of CO,  CO and O$_2$ adsorb and desorb on the surface\cite{ReuterScheffler:06}.  CO remains bonded upon adsorption whereas O$_2$ dissociates, two adjacent sites will become occupied by a single O molecule.
In this case the set of possible site states is $\mathcal{S}=\{\emptyset, \text{O}, \text{CO}\}$, which corresponds to an unoccupied site, a site occupied by O, and a site occupied by CO, respectively.

Supposing that we start with an \mbox{$N\times N$} surface, the master equation is formulated with a known transition matrix, $A_{ij}$, which prescribes the rate at which one particular system state transitions to another and may be written as
\begin{equation}
\frac{dP(\mathbf{S}_i)}{dt} = \sum_{j} A_{ji} P(\mathbf{S}_j) - A_{ij} P(\mathbf{S}_i),
\label{eqn:fullCME}
\end{equation}
where $P(\mathbf{S}_i)$ is the probability of the surface being in state $\mathbf{S}_i$, and a state $\mathbf{S}_i\in\mathcal{S}^{N\times N}$ is an $N\times N$ vector describing the state of each site (for example, refs. \onlinecite{Gillespie:92, Karlin, Jansen}).
In the discrete setting of surface reactions, each site may be in one of $|\mathcal{S}|$ states, where $|\cdot|$ represents the cardinality of a set.
The master equation yields a set of $|\mathcal{S}|^{N^2}$ ordinary differential equations (ODEs).
In the limit of $N\rightarrow \infty$, the kinetics are described by a denumerable but infinite dimensional ODE (so long as $|\mathcal{S}|>1$), an intractable problem that requires truncation along with periodic boundary conditions to become solvable.
Even for finite values of $N$, the size of state space is often intractably large for realistic choices of $N$, and thus instead of solving the master equation directly, a stochastic realization of the surface dynamics is often considered {\color{\red} by} using kinetic Monte Carlo algorithms (see for example, ref. \onlinecite{Jansen}).

In the present work, we restrict attention to translationally invariant systems and hypothesize that they exhibit a finite correlation length captured in an $m \times n$ subdomain we refer to as a tile within the overall $N \times N$ domain, ($m,n \ll N$). {\color{\red} As suggested by the notation, we will consider square lattices in the present work, however see no reason why the theory cannot be generalized to different lattice geometries.} Under these assumptions, a truncated BBGKY hierarchy for approximation of the master equation can be obtained based on the dynamics of tiles that capture correlation effects.   In essence the procedure is yet another phenomenological closure, but of increasing accuracy with increasing tile size. Mathematically, this corresponds to construction of a stochastic reduced model of size $m \times n$ to approximate the behavior of a larger $N \times N$ system. The formal analysis of the approximation error will be presented in follow-on work. Here we present the overall procedure and demonstrate the practical efficiency of the approach for surface CO catalysis.
Consider a rectangular tile of size $m\times n$ that covers $m\times n$ contiguous sites within the overall $N \times N$ surface, $m,n\ll N$ (see Figure 1 for an illustration of $1\time1$ and $2 \times 1$ tiles). Sites may be of various types (e.g., bridge or cus, cf. Fig. 1); let $\mathcal{T}$ denote the set of site types (distinct from $\mathcal{S}$, the set of site states). 
For some chosen $m \times n$ tile, let $\mathcal{T}'$ denote the set of possible tile types.
This is usually a small subset of $\mathcal{T}^{m \times n}$ fixed by the overall lattice construction.
For example, in a one-dimensional (1D) lattice with repeating $-ABBA-$ sites, the possible $3 \times 1$ tiles are $\mathcal{T}'=\{ABB, BBA, BAA, AAB\}$, a subset with four elements of the eight-element set $\mathcal{T}^{m \times n}$. It is necessary to distinguish tiles of the same type that have different neighbors. For example in the 1D lattice $-ABBBC-$, the are two variants of the $2 \times 1$ BB tile, one with $A,B$ neighbors, the other with $B,C$ neighbors. We therefore introduce $\Xi$ as the set of tile types with distinct positions in the lattice.

Having categorized each tile {\color{\red}$\xi\in\Xi$, we may denote the state of the tile, \sns{ in state} $\mathbf{s}\in\mathcal{S}^{m\times n}$, as \sns{(denoted} $\xi(\mathbf{s})$\sns{)}}, with underlying site types $\tau\in\mathcal{T}'${\color{\red}\sns{, w}. W}e may then assign a discrete PDF of finding tile $\xi$ in state $\mathbf{s}$, and denote this probability as $P_\xi(\mathbf{s})\equiv P(\xi(\mathbf{s}))$.
Due to the assumed translational invariance of the system, all tiles identified with tile type $\xi$ in the lattice are assumed to have identical probability distributions throughout time.
We seek to approximate these dynamics by assuming knowledge only up to a given \mbox{$(m\times n)$} tiling on the system, which will hold information of the PDF's $\{P_\xi\}_{\xi\in\Xi}$.
Each PDF has a domain of size $|\mathcal{S}|^{m\times n}$, there are $|\Xi|$ PDF's to track, and thus the goal is to reduce the large or infinite dimensional master equation (Equation \ref{eqn:fullCME}) to a $|\Xi|\times|\mathcal{S}|^{m\times n}$ dimensional set of ordinary differential equations.  Because $|\Xi|=|\mathcal{T}'|$ in many interesting cases (such as the (110) and (100) surfaces described above), the dimensionality of the ODE will often be equivalent to $|\mathcal{T}'|\times|\mathcal{S}|^{m\times n}$.

To achieve this approximation, we first note that given an arbitrary \mbox{$(m\times n)$} tile located on the $N\times N$ surface with site type geometry $\tau\in\mathcal{T}'$ and lattice position described by $\xi\in\Xi$, we may evaluate the dynamics of the discrete PDF over the state space on this tile based on the full description of the master equation (Equation \ref{eqn:fullCME}).
Below we will refer to `the tile $\xi$,' by which we mean an arbitrary choice from all the similar tiles $\xi$ from the full $N\times N$ system.
We reiterate that the reason we may choose any arbitrary $\xi$ is due to the assumed translational invariance of the system.
To approximate the dynamics of the PDF $P_\xi$ on a reduced system, we decompose the transition matrix $A$ into a sum of three matrices: a matrix that only changes site states within the tile $\xi$, $\tilde{A}^\xi$, a matrix that changes site states both within the tile and exterior to the tile, $\bar{A}^\xi$, and a matrix that does not change any of the site states within the tile, $\underline{A}^\xi$.
Thus we write
\begin{equation}
A=\tilde{A}^{\xi}+\bar{A}^{\xi}+\underline{A}^{\xi}.
\end{equation}
The second matrix, $\bar{A}^\xi$ is further decomposed by considering collections of sites that include an arbitrary number of exterior sites along with all of the sites in the tile $\xi$, and in which the site states of $\xi$ are specified.  Any one of these collections will be denoted $\mathbf{r}_\xi(\sigma)$ where $\mathbf{r}_\xi$ is a collection of sites that includes the sites in $\xi$, and $\sigma$ represents the states of each site in the collection $\mathbf{r}_\xi$.  We then define $\bar{A}^{\mathbf{r}_\xi(\sigma)}$ to be the matrix that encodes transitions such that (i) at least one site in the tile $\xi$ changes state, (ii) \emph{every} site exterior to the tile $\xi$ in $\mathbf{r}_\xi$ is changed, and (iii) \emph{no} site exterior to $\mathbf{r}_\xi$ changes state.  Summing over all possible choices of $\mathbf{r}_\xi(\sigma)$, both in selection of $\mathbf{r}_\xi$ and the choice of states $\sigma$, we then further decompose $\bar{A}$ to be
\begin{equation}
\bar{A}^{\xi} = \sum_{\mathbf{r}_\xi(\sigma)} \bar{A}^{\mathbf{r}_\xi(\sigma)}.
\end{equation}

Because we are only interested in the local events, we then average the rate of each transition type in the tile going from state $\mathbf{s}_i$ to state $\mathbf{s}_j$ (both states in $\mathcal{S}^{m\times n}$).  We describe these averaged transitions rates as
\begin{subequations}
\label{eqn:meanfield}
\begin{equation}
\tilde{a}^{\xi}_{ij} = \sum_{\mathbf{b}\in S(\Omega \backslash \xi)} \tilde{A}^{\xi}_{ij}(\mathbf{b}) P(\mathbf{b} | \xi(\mathbf{s}_i), \{P_{\xi'}\}_{\xi'\in\Xi}),\label{eqn:inmatrxgenreduct}
\end{equation}
\begin{equation}
\bar{a}^{\mathbf{r}_\xi(\sigma_i)}_{ij} = \sum_{\mathbf{b}\in S(\Omega \backslash \mathbf{r}_\xi)} \bar{A}^{\mathbf{r}_\xi(\sigma_i)}_{ij}(\mathbf{b})P(\mathbf{b} | \mathbf{r}_\xi(\sigma_i), \{P_{\xi'}\}_{\xi'\in\Xi}),\label{eqn:outmatrxgenreduct}
\end{equation}
\end{subequations}
where $S(\Theta)$ is the set of all possible states of the collection of sites $\Theta$, $\xi$ is the tile that may also be thought of as a set of sites on the surface $\Omega$ that comprise the tile, and $\sigma_i$ is a choice of $\sigma$ that is constrained so that the states on the tile are described by $\mathbf{s}_i$.
The matrix elements $B^{\mathbf{q}}_{ij}(\mathbf{b})$ in the interior of each sum are the elements of the transition matrix that begin in state $\mathbf{s}_i$ on the subset $\xi\subset\mathbf{q}$ and finish in state $\mathbf{s}_j$ on $\xi$, and all exterior states in $\Omega\backslash\mathbf{q}$ remain fixed in state $\mathbf{b}$.  The last condition will be trivially satisfied based on the definitions of the matrix decomposition above.  Finally the sum is weighted by the conditional probability that the system will be in state $\mathbf{b}$ given that we know either the state of the tile $\xi(\mathbf{s}_i)$ or the state of tile along with the external sites specified by $\mathbf{r}_\xi(\sigma_i)$ and the discrete PDF's of all tiles.

We then may define new transition matrices $\tilde{a}^{\xi}$ and $\bar{a}^{\mathbf{r}_\xi}$, all of which have dimension $|\mathcal{S}^{m\times n}|\times|\mathcal{S}^{m\times n}|$. Taken together, these matrices represent a mean field theory that accounts only for the sites that change within the tilings, averaging the influences of the system states that do not change with a given transition; we note that in general, the reduced matrices will depend non-trivially on the PDF's of the tiling, however in the current work we will primarily focus on systems that have transition rates independent of the state of sites that do not change states.  Therefore, having a fixed $i,j$ and $\mathbf{q}$, we assume that $B^{\mathbf{q}}_{ij}(\mathbf{b}_1)=B^{\mathbf{q}}_{ij}(\mathbf{b}_2)$ for all $\mathbf{b}_1,\mathbf{b}_2\in S(\Omega\backslash\mathbf{q})$, which means that we may arbitrarily assign any of these elements to the reduced matrix:
\begin{subequations}
\label{eqn:simpmeanfield}
\begin{equation}
\tilde{a}^{\xi}_{ij} = \tilde{A}^{\xi}_{ij}(\mathbf{b}), \quad \forall \mathbf{b}\in S(\Omega \backslash \xi),\label{eqn:inmatrxreduct}
\end{equation}
\begin{equation}
\bar{a}^{\mathbf{r}_\xi(\sigma_i)}_{ij} = \bar{A}^{\mathbf{r}_\xi(\sigma_i)}_{ij}(\mathbf{b}), \quad \forall \mathbf{b}\in S(\Omega \backslash  \mathbf{r}_\xi).\label{eqn:outmatrxreduct}
\end{equation}
\end{subequations}

Having defined the transition rates on tiles, the master equation approximation is
\begin{eqnarray}
&&\frac{dP_\xi(\mathbf{s}_i)}{dt} =~~\sum_{j=1}^{|\mathcal{S}|^{m n}} \Big(\tilde{a}_{ji} P_\xi(\mathbf{s}_j) - \tilde{a}_{ij} P_\xi(\mathbf{s}_i)\Big) \nonumber\\
&&+\sum_{j=1}^{|\mathcal{S}|^{m n}} \sum_{\mathbf{r}_\xi(\sigma_j)} \bar{a}_{ji}^{\mathbf{r}_\xi(\sigma_j)} P_\xi(\mathbf{s}_j) P(\mathbf{r}_\xi(\sigma_j)| \xi(\mathbf{s}_{j}), \{P_{\xi'}\}_{\xi'\in\Xi})\nonumber \\
&&-\sum_{j=1}^{|\mathcal{S}|^{m n}} \sum_{\mathbf{r}_\xi(\sigma_i)} \bar{a}_{ij}^{\mathbf{r}_\xi(\sigma_i)} P_\xi(\mathbf{s}_i)P(\mathbf{r}_\xi(\sigma_i)|\xi(\mathbf{s}_i), \{P_{\xi'}\}_{\xi'\in\Xi}).\label{eqn:mastertruncate}
\end{eqnarray}
The first summation on the right hand side represents tile state changes that occur entirely within tile.  The second and third summation capture state changes dependent on neighboring tiles being in a specific state.

To close the set of equations given in equation \ref{eqn:mastertruncate}, the conditional probabilities 
\begin{equation}
P(\mathbf{r}_\xi(\sigma)| \xi(\mathbf{s}), \{P_{\xi'}\}_{\xi'\in\Xi}), \label{eqn:condprob}
\end{equation}
 must be computed.  These are the probabilities that the neighboring sites of $r_{\xi}$ are
in state $\sigma$ when the tile is in state $\mathbf{s}$. 
In this work we consider square lattices in which only a single site exterior to the tile $\xi$ changes state upon a reaction.  We note that the second assumption is common as many model reactions effect either one or two sites, but often no more.  With this said, we note that it is possible to formally determine equation \ref{eqn:condprob} with the both assumptions relaxed, however such cases are beyond the scope of the current work.  Next we say that if there is no tile that covers both the exterior changed tile and the interior changed tile, then the probability of finding the exterior tile in a given state is independent of the interior state.  Along with the consistency criteria presented below (see section \ref{sec:constraint}), this will provide a robust method for determining the probability of finding the exterior site in the desired state.

\begin{figure}
\includegraphics[width=4cm]{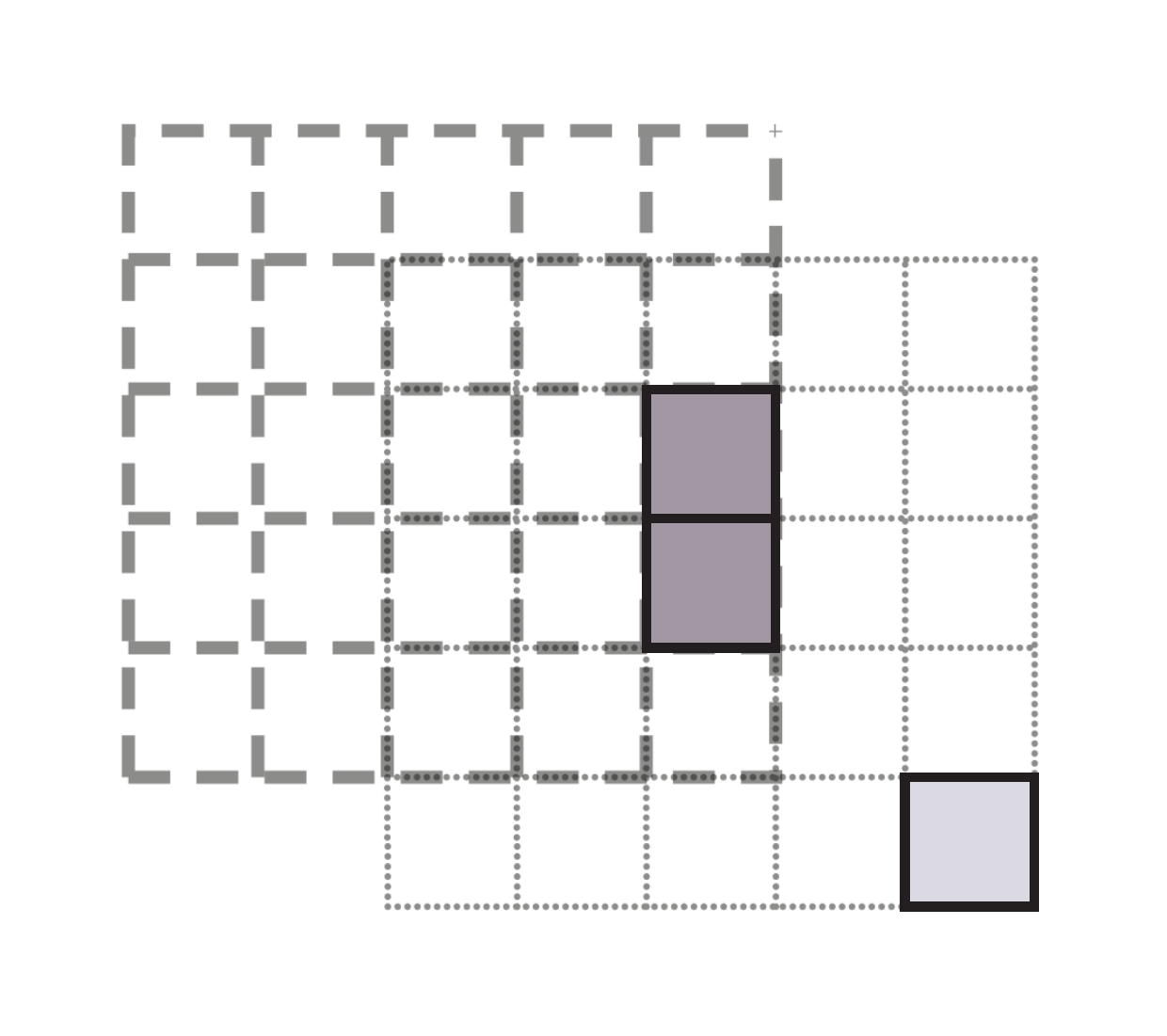}
\includegraphics[width=4cm]{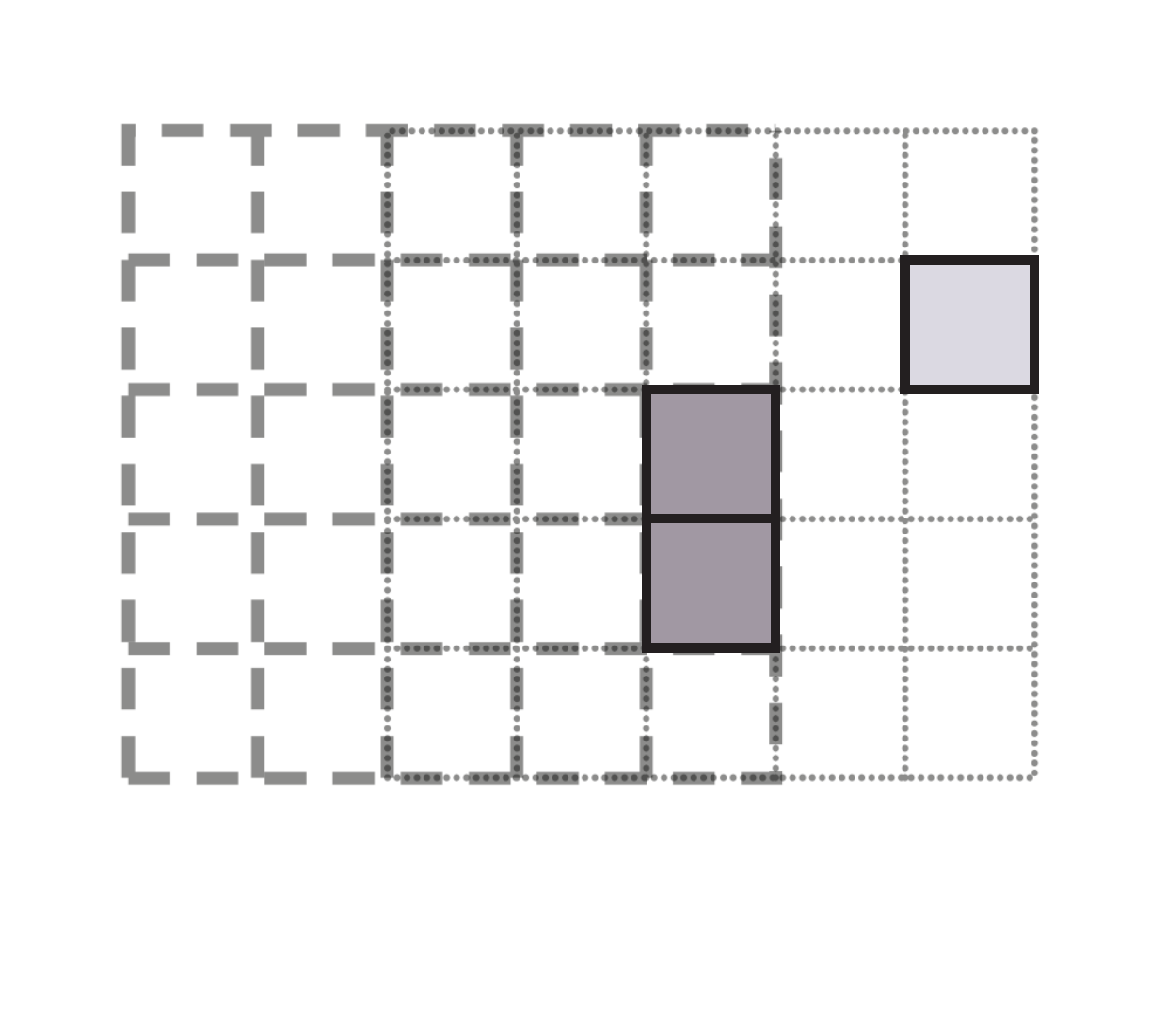}
\caption{We demonstrate two examples of how to reconstruct conditional probabilities where the transition will occur over both interior and exterior states to an original tile.  The original tile is represented with thick dashed lines, the interior sites that will transition are represented by the sites that are darkly shaded with black outline, and the exterior site that will transition is shaded in light grey with black outline.   The exterior site must be in state $s_{ext}$ for the transition to be able to occur.  We integrate over all states of a secondary overlapping tile, represented with thin solid grey lines, excluding the exterior tile site that will change state.}
\label{fig:closure}
\end{figure}

As an example of the calculation of the conditional probability (equation \ref{eqn:condprob}), consider the $5 \times 5$ tile shown in Figure \ref{fig:closure} along with a reaction that changes the states of the two sites (5,2), (5,3) within the tile, but requires a neighbor site outside the tile to be in a specific state. We must determine the probability of finding the overall initial state of both tile and neighboring sites that allows for the reaction to occur.  This is computed by summation of the probability of all partially overlapping tiles that
have consistent states.

To obtain a precise computational statement of the required conditional
probability, consider \ the tile $\xi (\mathbf{s})$, a tile of type $\xi \in
\Xi$, in initial state $\mathbf{s} \in \mathcal{S}^{m \times n}$. Let $\xi
(s_{x, y})$ denote both {\color{\red}\sns{type}the} site type and state at position $(x, y)$ within
the tile, $1 \leqslant x \leqslant m$, $1 \leqslant y \leqslant n$. Consider a
state transition at site $(x, y)$ that requires a neighbor site $(x + u, y +
v)$ situated outside the tile to be in state $s_{\tmop{ext}}$. Let $\Xi_{(a,
b)} (\xi (\mathbf{s}))$ denote the subset of tiles that have the same site
types and states as tile $\xi (\mathbf{s})$ in the overlap region after
translation by $(a, b)$
\[ \Xi_{(a, b)} (\xi (\mathbf{s})) = \{ \eta (\mathbf{r}) \nocomma | \eta
   (\mathbf{r}_{(c, d)}) = \xi (\tmmathbf{s}_{(a + c, b + d)}), 1 \leqslant a,
   a + c \leqslant m, 1 \leqslant b, b + d \leqslant n \} . \]
The probability of a translated tile to have the correct overlap site types
and states is
\[ A = \sum_{\eta (\mathbf{r}) \in \Xi_{(a, b)} (\xi (\mathbf{s}))} P (\eta
   (\mathbf{r})) . \]
Within $\Xi_{(a, b)} (\xi (\mathbf{s}))$ there exists a further subset
$\Xi^{(u, v)}_{(a, b)} (\xi (\mathbf{s}))$ that has the required state
$s_{\tmop{ext}}$ at position $(x + u, y + v)$
\[ \Xi^{(u, v)}_{(a, b)} (\xi (\mathbf{s})) = \Xi_{(a, b)} (\xi (\mathbf{s}))
   \cap \{ \eta (\mathbf{r}) \nocomma, \eta (\mathbf{r}_{(x + u - a, y + v -
   b)}) = s_{\tmop{ext}} \} . \]
The conditional probability from (7) is given by
\begin{equation} P (\nobracket \mathbf{r}_{\xi} (\sigma) | \xi (\tmmathbf{s}_i)) =
   \frac{1}{A} \sum_{\eta (\mathbf{r}) \in \Xi^{(u, v)}_{(a, b)} (\xi
   (\mathbf{s}))} P (\eta (\mathbf{r})),  \label{eqn:closure} \end{equation}
with numerator simply expression the further restriction on the required
$s_{\tmop{ext}}$ state at $(x + u, y + v)$.

This methodology provides a robust method for closing the approximated dynamics of the master equation with the reduced tiling system provided that only one site exterior to a given tile changes with arbitrarily many interior sites based on all given transitions.  As we have mentioned above, it is also straight forward to generalize this framework to cases in which multiple interior sites, or multiple exterior sites transition, however such reactions are not considered in the present work and we forgo this discussion presently.   

Having established a reduced system with corresponding equations for arbitrary rectangular tilings, we next mention some simple ways in which we can take advantage of rotational symmetries on the lattice, and then introduce a generalization to a mixed tiling scheme.   Following this exposition, to ensure both the ODE written in equation \ref{eqn:mastertruncate}, and the generalization presented below are self-consistent, we state two consistency criteria.  From here, we present a simple example, and then {\color{\red}several\sns{the}} concrete example{\color{\red}s} of surface catalysis, which will be used below in numerical tests.

\subsection{Rotational symmetry and mixed tilings}
Up to this point, we have discussed tilings with fixed orientations.  Certain lattices, however, contain symmetries that may be used to obtain higher range spatial correlations without increasing the dimension of the set of corresponding equations.  In the (100) crystal above, for example, there are four rotational symmetries of the lattice found by rotation of $\pi/2$ radians.  Thus we expect that for each $(m\times n)$ tile, there is a corresponding $(n\times m)$ tile with an equivalent PDF over site states under rotation.  Therefore, in reconstructing the conditional probabilities used in conjunction with $\bar{a}$, we assume that we have both $(m\times n)$ and  $(n\times m)$ tiling types.  In the case that $m\neq n$, this may lead to more detail in spatial correlations along both lattice directions, and thus a more accurate method with no additional cost.

We may also consider the (110) surface described above that only has two rotational symmetries found by rotation of $\pi$ radians.  In this case, considering $(m\times1)$ tiles would provide us with spatial correlations along cus-cus or bridge-bridge sites but we would be assuming independence in the cus-bridge direction.  Similarly a $(1\times n)$ tiling would retain spatially correlated data in the bridge-cus direction, but not along bridge-bridge or cus-cus networks.  To get around this we could consider $(m\times n)$ tiling, however the number of equations would grow exponentially, from an $|\Xi_{(m\times 1)}|\times|\mathcal{S}|^m$ to a 
$|\Xi_{(m\times n)}|\times|\mathcal{S}|^{(m\times n)}$ dimensional ODE, where $\Xi_{(m\times n)}$ is the tile set when considering a $(m\times n)$ tiling.  In the case of the (110) surface $|\Xi_{(m\times 1)}| = 2$, and $|\Xi_{(1\times n)}|$ and $|\Xi_{(m\times n)}|$ are equal to $1$ if $n$ is even and $2$ if $n$ is odd.
Instead we may consider mixed tilings, which is to say we consider the system constructed when considering the PDF's on $(m\times 1)$ and $(1\times n)$ tiles, which causes the system to grow from a $|\Xi_{(m\times 1)}|\times|\mathcal{S}|^m$ to a $|\Xi_{(m\times 1)}\cup\Xi_{(1\times n)}|\times(|\mathcal{S}|^m+|\mathcal{S}|^n)$ dimensional {ODE} (so long as $m,n>1$).  
To close the conditional probabilities we utilize an identical method to that listed above, however note that  in some cases we will be able to choose either an $m\times n$ or $n\times m$ overlapping tile that satisfies the above criteria.  To make the choice unique we will first require that the chosen tile containing both the interior and exterior sites that will transition, contain a maximal number of overlapping sites.  In the case that this choice is not unique, we require the sum of the radial distance between the overlapping tiles and the interior transitioning tiles to be minimal, where we define the radial distance in the usual sense (i.e. the distance between site $ij$ and $lk$ is $\sqrt{(i-l)^2+(j-k)^2}$).  We conjecture but are not certain that this criteria will provide a unique tiling choice, but mention that it does in all the cases we have considered below.

With all of the methods listed above, we must ensure that several constraints are met to ensure the system is consistent.  We expose these constraints in the following subsection.

\subsection{Consistency and constraints on tiling dynamics}\label{sec:constraint}
Care must be taken to ensure that certain constraints are obeyed by any of the resulting reduced systems described above.
For one, a system must be initialized to have normalized PDF's over each tile.
Analytically, the PDF's will trivially be normalized throughout all time, since what is added to one state is taken away from another.
Next all lower dimensional projections must be well defined.
By this we mean that lower dimensional projections must agree between (i) different tiles $\xi\in\Xi$ and (ii) within a given tiling.  
For example, suppose we have a mixed $(2\times1)$ and $(1\times2)$ tiling on a (110) surface as described above, suppose we have a bridge-bridge tile, $\xi$ and a bridge-cus tile $\xi'$.  We note that in the $(1\times1)$ tiling system all bridge and cus sites are identical on the lattice and thus the probability of finding a bridge site in a particular state must be the same no mater how it is determined.  This means that we must have
\begin{subequations}
\label{eqn:constraints}
\begin{equation}
\sum_{s_i\in \mathcal{S}} P_{\xi}((s,s_i)) = \sum_{s_i\in \mathcal{S}} P_{\xi'}((s,s_i)), \forall s\in\mathcal{S},
\end{equation}
\begin{equation}
\sum_{s_i\in \mathcal{S}} P_{\xi}((s,s_i)) = \sum_{s_i\in \mathcal{S}} P_{\xi}((s_i,s)), \forall s\in\mathcal{S}.
\end{equation}
\end{subequations}
We conjecture that systems will be well behaved for rectangular $m\times n$ tiling systems and for mixed tiling schemes.  We note that we have numerically verified that all systems mentioned below satisfy the above criteria for a variety of test cases and projections.  From this point on we will assume all tiling systems are mixed, i.e. saying that we are working with a $(2\times 1)$ tiling system will mean that we are working with a mixed $(2\times1)$ and $(1\times2)$ tiling system.

In general, Equation \ref{eqn:mastertruncate} may be difficult to work with explicitly, however we will show that in several relevant problems it simplifies to a corresponding hierarchy of ODEs that can be easily coded into a computational algorithm.

\subsection{A simple example}\label{sec:simpeg}
We illustrate the above concepts with a simple example.
Suppose that $\Omega$ is comprised of two sites each of which is identical, and each of which can be in states $\mathcal{S}=\{0,1\}$.
The state space is then $(00, 01, 10, 11)^T$.
The transition matrix is chosen as
\begin{equation}
A =
\left( \begin{array}{cccc}
-2 r_1 & 0 & 0 & 0 \\
r_1 & -r_2 & 0 & 0 \\
r_1 & 0 & -r_2 & 0\\
0 & r_2 & r_2 & 0 \end{array} \right),
\end{equation}
which states that a site in state 0 can transition to state 1, but not the other way around, and that the rate with which this happens depends on the state of the other site.  The four dimensional system may be written as
\begin{eqnarray}
\dot{P}_{00}(t) =& -2r_1 P_{00},  \quad &\dot{P}_{01}(t) = r_1 P_{00} -  r_2 P_{01},\nonumber \\
\dot{P}_{10}(t) =& r_1 P_{00} -  r_2 P_{10},  \quad  &\dot{P}_{11}(t) = r_2 P_{01}+r_2 P_{10}.
\end{eqnarray}
Suppose we choose a $(1\times1)$ tiling and wish to determine the reduced dynamics. There is only one type of tile as both sites share identical dynamics. We find that $\bar{A}=0$ (and hence $\bar{a}=0$) as there are no reactions effecting both sites and
\begin{equation}
\tilde{a} = \left( \begin{array}{cc}
-r_1P_0-r_2 P_1 & 0  \\
r_1P_0+r_2 P_1 & 0
\end{array} \right),
\end{equation}
corresponding to the state space $(0,1)^T$, which leads to dynamics given by
\begin{equation}
\dot{P}_{0}(t) = -(r_1P_0+r_2 P_1)P_0,  \quad \dot{P}_{1}(t) = (r_1P_0+r_2 P_1)P_0.
\end{equation}

\subsection{A concrete example on an idealized (110) surface}

We next demonstrate how to construct approximated systems with a more realistic example.
To do this, we consider the (110) surface made up of bridge and cus sites (see Figure \ref{fig:surfaces}).
Each bridge site is connected to two other bridge sites and two cus sites.
We suppose that we know a collection of surface transitions and each of their rates.
In the following example we consider approximate dynamics of CO oxidation for $(1\times1)$ and $(2\times1)$ tilings with the collection of possible events that can generate state changes on the surface, given as
\begin{subequations}
\label{eqn:surfacereactions}
\begin{eqnarray}
\text{CO}_g + \emptyset_i \rightleftharpoons& \text{CO}_i  &\text{(ads/des),}\label{eqn:siter}\\ %
\text{O$_2$}_g + \emptyset_i + \emptyset_j \rightleftharpoons& \text{O}_i + \text{O}_j  &\text{(ads/des),}\label{eqn:pairri}\\
\text{CO}_i + \text{O}_j \rightarrow& \text{CO$_2$}_g + \emptyset_i + \emptyset_j  &\text{(surf rxn),}
\label{eqn:pairrf}
\end{eqnarray}
\end{subequations}
where the subscripts $i,j$ represent the site types being occupied and the subscript $g$ represents molecules in the gas phase disconnected from the surface.
The only additional constraint is that sites must be adjacent on the lattice for transitions involving two sites to occur.  The classification ads/des denotes adsorption and desorption events (depending on the reaction direction) and {\color{\red}`}surf rxn{\color{\red}'} denotes surface reactions.
We classify the transitions as site transitions (e.g. Equation \ref{eqn:siter}) or pair transitions (e.g. Equation\ref{eqn:pairri}).
We may also define diffusion events as pair transitions that appear as
\begin{equation*}
\text{CO}_i + \emptyset_j \rightarrow \text{CO}_j + \emptyset_i,
\end{equation*}
but we do not consider these events in the present work.  { The reason for this omission is four-fold: (1) near steady state the systems we examine have very low probability of being in the empty state and thus diffusion in these regimes will not heavily influence the system, (2) many previous studies that have explored this work have also left out the consideration of diffusion (see for example refs \onlinecite{Mai:93,Temeletal:07}) and thus omitting diffusion will make our work more comparable to  these studies, (3) we have run simulations where we have included diffusion (but have omitted the results from this presentation) and find that the system dynamics are nearly identical to the case without diffusion, and (4) we note that diffusion tends to mix systems and thus smaller elements of the hierarchy may lead to accurate results in cases where diffusion is important; thus we expect that showing that the hierarchy is still effective when diffusion is not relevant should lead to a more stringent verification of our method.}
There are two possible site types ($\mathcal{T}=\{$bridge, cus$\}$), and each may be in one of the three possible states, $\mathcal{S}=\{\emptyset$, O, CO$\}$.

\subsubsection{$(1\times1)$ tilings}
We begin with a $(1\times1)$ tiling and show that this tiling leads to {\color{\red}\sns{an identical} a similar} set of ODEs to the PK model described in ref. \onlinecite{Temeletal:07}.
There are two possible tilings, $\xi$, representing a covering of each site type.
We shall describe the set of tilings as $\Xi = \{$[b], [c]$\}$ for bridge and cus sites respectively.
We next wish to determine an approximate master equation describing the probability of finding each particular site in a certain state.
Following the formalism of the previous section, the only possible interior transition in each tile is the site transition (Equation \ref{eqn:siter}), and the other transitions enter by the second and third terms in Equation \ref{eqn:mastertruncate}.
The mixed interior/exterior terms are closed via Equation \ref{eqn:closure}.
Each transition occurs stochastically with an exponential distribution in time; this implies that taking a time step of $dt$ a single event occurs with probability proportional to $dt$.
Multiple events are considered independent and thus occur with probability proportional to $dt^2$.
In the limit as $dt\rightarrow0$, these terms will vanish leaving a simple transition matrix that only describes events that generate state changes on a site or two adjacent sites via the site and pair transitions listed above, respectively, with transition rates that are assumed to be independent of the sites that are not effected by the transition.
This observation greatly reduces the space of possible choices $\mathbf{r}_\xi$, and allows us to use the reduction presented in equations \ref{eqn:inmatrxreduct} and \ref{eqn:outmatrxreduct}.
We note further that there are multiple choices for $\mathbf{r}_\xi$ that will lead to identical reactions.
For example given a $(1\times 1)$ tile covering a single bridge site, bridge-bridge pair transitions can effect the bridge tile either through the left or right bridge site.
Based on the symmetry of the system, each transition rate will be identical, and thus instead of accounting for these distinct choices for $\mathbf{r}_\xi$, we combine and weight them with a weight function $w(\xi'|\xi)$ which describes the number of adjacent tiles $\xi'$ given that we are at the tile $\xi$.  This weight function has arisen naturally by summing over $\mathbf{r}_\xi(\sigma)$ and will appear in the $\bar{a}^\xi$ terms of equation \ref{eqn:mastertruncate}.
We may write the resulting six-dimensional {set of equations} explicitly as
{\color{\red}\begin{eqnarray}
&&\frac{dP_\xi(\mathbf{s})}{dt} = \sum_{\mathbf{t}\in\mathcal{S}} k_{\xi(\mathbf{t})\rightarrow\xi(\mathbf{s})} P_\xi(\mathbf{t}) - k_{\xi(\mathbf{s})\rightarrow\xi(\mathbf{t})} P_\xi(\mathbf{s}),\nonumber\\
&&- \sum_{\substack{\xi'\in \Xi\\ \mathbf{q},\mathbf{r},\mathbf{t}\in\mathcal{S}}} w(\xi'|\xi) \left(k_{\xi(\mathbf{s})\xi'(\mathbf{t})\rightarrow\xi(\mathbf{q})\xi'(\mathbf{r})}P_\xi(\mathbf{s}) P_{\xi'}(\mathbf{t})\right)\nonumber\\
&&+ \sum_{\substack{\xi'\in \Xi, \\ \mathbf{q},\mathbf{r},\mathbf{t}\in\mathcal{S}}} w(\xi'|\xi) \left(k_{\xi(\mathbf{q})\xi'(\mathbf{r})\rightarrow\xi(\mathbf{s})\xi'(\mathbf{t})}P_\xi(\mathbf{q}) P_{\xi'}(\mathbf{r})\right)\label{eqn:1x1tiling},
\end{eqnarray}
where $\mathbf{s}\in\mathcal{S}$.  The reaction speeds $k_{\xi(\mathbf{s})\rightarrow\xi(\mathbf{t})}$ and $k_{\xi(\mathbf{q})\xi'(\mathbf{r})\rightarrow\xi(\mathbf{s})\xi'(\mathbf{t})}$ are set to zero if the reactions corresponding to the transitions within the subscript do not occur (it is assumed in this notation that tile $\xi$ is adjacent to tile $\xi'$).}
The weight function $w(\xi'|\xi)$ is computed from the geometry of the system.
In the current geometry, each cus has two bridge and two cus neighbors, so that $w([b]|[c]) = w([c]|[c]) = 2$.
{\color{\red}\sns{A similar calculation is performed over all pairings.}}
This {set of equations} {\color{\red} \sns{is identical to} takes on the same form as} the PK model found in ref. \onlinecite{Temeletal:07}.

\subsubsection{$(2\times1)$ tilings}
\label{subsec:2x1}
We continue with a $(2\times1)$ tiling.
We note that in the given geometry there is a one-to-one mapping between $\xi$ and it{\color{\red}\sns{'}}s site types $\tau=(\tau_1,\tau_2)\in\mathcal{T}'$, and thus denote $\xi$ by it{\color{\red}\sns{'}}s pair of site types.
In this case, there are three possible tilings (up to symmetry and assuming a mixed tiling scheme) based on the given geometry, given as $\Xi=\mathcal{T}'=\{$[b,b], [b,c], [c,c]$\}$.
We wish to determine an approximation of the master equation describing the probability of finding each particular pair in a certain state.
Note that all of the listed transitions may occur within the interior of the listed tiles and thus contribute to $\tilde{a}$.
In the current setting we now have a $|\Xi|\times|\mathcal{S}|^{2\times1}=3\times3^2=27$ dimensional {set of equations}.
Let $p_{lm}^{ij}$ denote the probability of finding til{\color{\red}e\sns{ing} $\xi$ with site types} $[i,j]$ in state $\mathbf{s}=(l,m)${\color{\red}, and $\xi_{ij}((l,m))$ denote the statement that tile $\xi$ with site types $[i,j]$ has the site of type $i$ in state $l$ and the site of type $j$ in state $m$}.
{\color{\red}\sns{We demonstrate the {equation} construction by writing the dynamics for {\color{\red}\sns{$p_{\emptyset\emptyset}^{ij}$}$p_{lm}^{ij}$}.}}  We again take advantage of the fact that the transition matrix will only account for local transitions, and that multiple choices of $\mathbf{r}_\xi$ will lead to identical contributions and so again introduce weight functions. Using Equation \ref{eqn:closure} to reconstruct the neighbor probabilities.   We obtain the {equations}
{\color{\red}
\begin{eqnarray}
&&\frac{{\color{\red}d}p_{lm}^{ij}}{dt} = 
\sum_{n\in\mathcal{S}}\Big(k_{n_i\rightarrow l_i}p^{ij}_{nm} +k_{n_j\rightarrow m_i}p^{ij}_{ln} \nonumber\\
&&~~-k_{l_i\rightarrow n_i}p^{ij}_{lm} -k_{m_j\rightarrow n_i}p^{ij}_{lm} \Big)\nonumber\\
&&~~+\sum_{n,o\in\mathcal{S}}k_{\xi_{ij}((n,o))\rightarrow\xi_{ij}((l,m))}p_{no}^{ij} - k_{\xi_{ij}((l,m))\rightarrow\xi_{ij}((n,o))}p_{lm}^{ij}\nonumber\\
&&~~+\sum_{\substack{\iota \text{ s.t. } \\ [\iota,i]\vee[i,\iota]\in \mathcal{T}', \\ n,o,q\in\mathcal{S}}}w([\iota,i] | [i,j]) k_{\xi_{i\iota}((o,q)) \rightarrow \xi_{i\iota}((l,n))}p_{om}^{ij}\frac{p_{oq}^{i\iota}}{\sum_{r\in\mathcal{S}} p_{or}^{i\iota}}\nonumber\\
&&~~-\sum_{\substack{\iota \text{ s.t. } \\ [\iota,i]\vee[i,\iota]\in \mathcal{T}', \\ n,o,q\in\mathcal{S}}}w([\iota,i] | [i,j]) k_{\xi_{i\iota}((l,n)) \rightarrow \xi_{i\iota}((o,q))}p_{lm}^{ij}\frac{p_{ln}^{i\iota}}{\sum_{r\in\mathcal{S}} p_{lr}^{i\iota}}\nonumber\\
&&~~+\sum_{\substack{\iota \text{ s.t. } \\ [\iota,j]\vee[j,\iota]\in \mathcal{T}', \\ n,o,q\in\mathcal{S}}}w([\iota,j] | [j,i]) k_{\xi_{\iota j}((o,q)) \rightarrow \xi_{\iota j}((n,m))}p_{lq}^{ij}\frac{p_{oq}^{\iota j}}{\sum_{r\in\mathcal{S}} p_{rq}^{\iota j}}\nonumber\\
&&-\sum_{\substack{\iota \text{ s.t. } \\ [\iota,j]\vee[j,\iota]\in \mathcal{T}', \\ n,o,q\in\mathcal{S}}}w([\iota,j] | [j,i]) k_{\xi_{\iota j}((n,m)) \rightarrow \xi_{\iota j}((o,q))}\frac{p_{lm}^{ij}p_{nm}^{\iota j}}{\sum_{r\in\mathcal{S}} p_{rm}^{\iota j}},
%
\end{eqnarray}}
where $w([\iota,i] | [i,j])$ is the weight function which gives the number of additional neighbors of type $\iota$ to site type $i$, given that we have already included a neighbor of type $j$.
For example, $w([b,b] | [b,b]) = 1$ and $w([c,b] | [b,b]) = 2$.  {\color{\red} The reaction rates effecting a single site are described by $k_{n_i\rightarrow l_i}$ denoting the rate at which a site of type $i$ transitions from state $n$ to state $l$.  Pair reactions rates are written similarly to the $(1\times1)$ case, however we now have a different way to describe the initial state as the initial states are covered by a single tile.}
The first sum represents {\color{\red} reactions that effect individual sites (i.e. adsorption/desorption of CO).  The second sum represents pair reactions (i.e. CO oxidation and adsorption/desorption of O$_2$) that occur on the interior of the tile.  The third and fourth sum represent pair reactions effect the first site of type $i$ from transition toward and away from state $l$, and fifth and sixth sums represent pair reactions that effect the second site of type $j$ from transitioning toward and away from state $m$.\sns{ site of neighboring pairs that share site $i$ in state $\emptyset$, with second site type $\iota$ in state $\emptyset$; in other words this term represents oxygen adsorption that effects sites both inside and outside the tile.  The remaining terms are analogous but for oxygen desorption (additive terms) and the other site of type $j$ (second and fourth terms).}}

Higher-order tilings, such as $(n\times1)$ and $(n\times n)$ tilings may be constructed similarly.  Note that in these higher-dimensional cases, the tiles will overlap on a greater number of sites and thus the conditional probability will account longer range spatial correlations.

\subsection{Algorithm construction}
The steps listed above may be generalized to a computational algorithm. However the weighting functions will change for each type of tiling. Given a lattice we may calculate the neighbor and conditional neighbor probabilities, we may construct the list of tiles $\xi$ along with their site types $\mathcal{T}'$ and position on the lattice, and we may construct a list of state transitions $\mathcal{R}$ along with the transition rates which may be looped through and added in accordance with the above procedures.  
We construct this algorithm and test the convergence of the approximated master equation (Equation \ref{eqn:mastertruncate}) in the proceeding section.  { We have begun to develop a code base for this algorithm and an example is available at https://github.com/gjherschlag/mixedtiling\_eg}.

\section{Numerical examples}
\subsection{CO oxidation on simplified (110) surfaces}
We continue with the example above of oxidation of CO on the (110) surface, however we simplify the model in two ways.  First, we treat the cus sites as being inactive which reduces the system to a one dimensional lattice.  Next, we assume there is no difference between cus and bridge sites and set the transition rates accordingly.  For both cases, we then have $\mathcal{S} = \{\emptyset,\text{O},\text{CO}\}$ and $\mathcal{T} = \{a\}$ along with the transitions listed in Equations \ref{eqn:siter}-\ref{eqn:pairrf}.  After analyzing these two test cases, we will then examine the system with differentiated bridge and cus sites and use the realistic parameters found in ref. \onlinecite{ReuterScheffler:06,Temeletal:07}.

For the test cases, we choose test parameters with ratios that are similar (in order of magnitude) to parameters found for the realistic system.  In non-dimensional units of time, we set $k_{\text{OO}\rightarrow\emptyset\emptyset} = 1.0, k_{\emptyset\emptyset\rightarrow\text{OO}} = 10^{-3}, k_{\text{COO}\rightarrow\emptyset\emptyset}=0.1, k_{\text{CO}\rightarrow\emptyset}=10^{-2}$, and vary $k_{\emptyset\rightarrow\text{CO}}\in[0,1]$.  { We note that in all parameter regimes the probability of finding a site in the empty state is close to zero for both KMC simulation and all examined levels of the hierarchy.  Because of this we report only the probability of the coverage of CO, $P(CO)$, as the oxygen coverage can be approximated by $P(O)\approx 1-P(CO)$ at steady state.}

\subsubsection{Inactive cus sites}
For the single dimensional (1D) lattice, we compare $(n\times1)$ tiling approximations of the master equation for $n\in\{1,2,3,4,5\}$.
We then compare the approximated master equation with a reaction-first KMC simulations on a $1024\times1$ periodic lattice.  A $(n\times1)$ tiling approximation makes up a $3^n$ dimensional {ODE}.  The ODE's are integrated using the LSODA routine wrapped in the scipy package for python \cite{Scipy}.  Details of the KMC simulation can be found in ref. \onlinecite{Jansen}, however we note that we use a GPU to accelerate the computation of the overall system reaction rate.

\begin{figure}
\includegraphics[width=9cm]{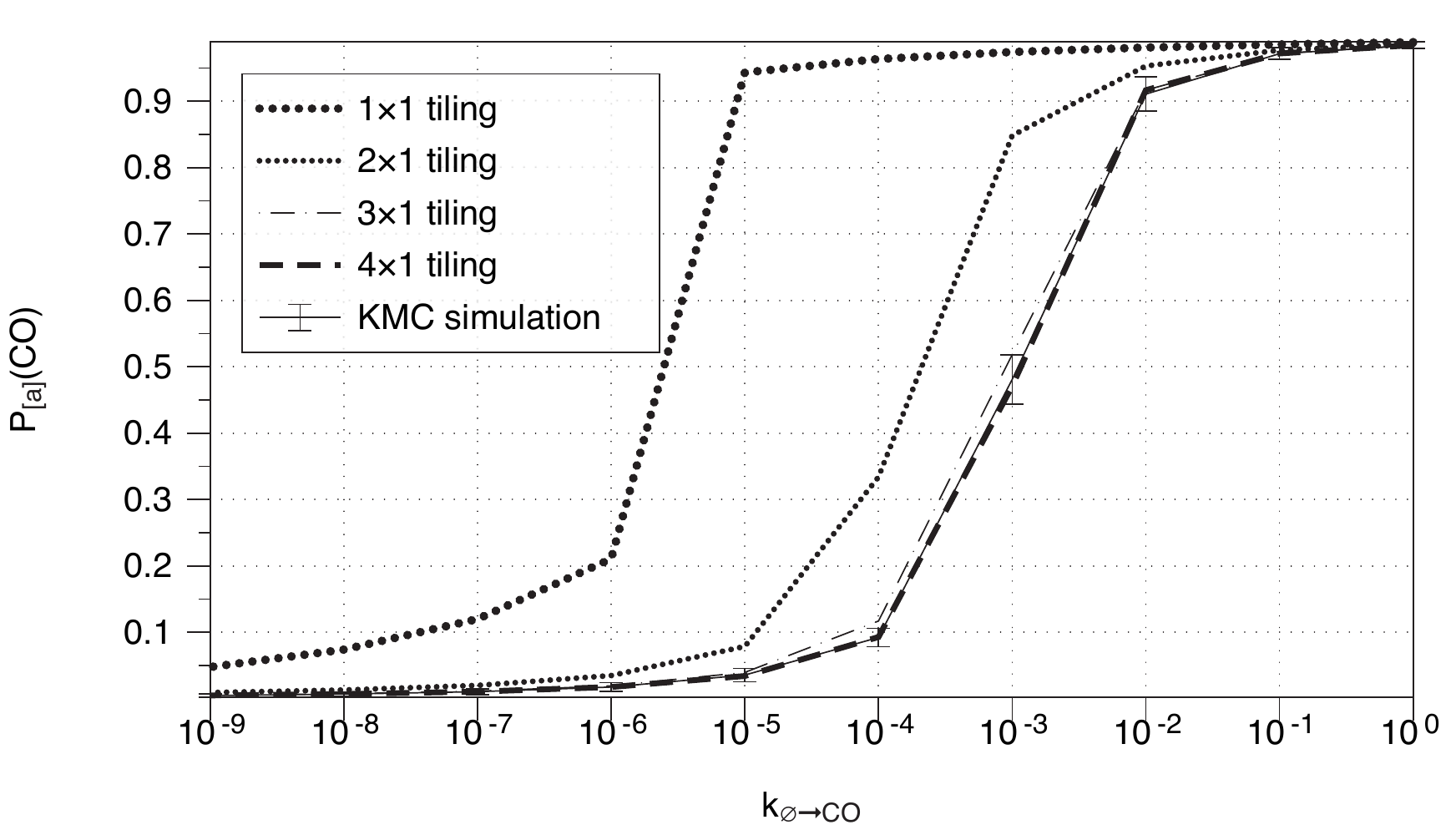}
\caption{For the 1D system, steady states values of the probability of CO on a single site are shown as the CO adsorption rate is increased for four tiling types and KMC results.  The $3\times1$ tiling approximation yields a solution that is within twice the standard deviation of the KMC mean statistics.  The $4\times1$ tiling approximation lies within a quarter of a standard deviation of the KMC mean statistics.}
\label{fig:ss1d}
\end{figure}

\begin{figure}
\mbox{\includegraphics[width=5cm]{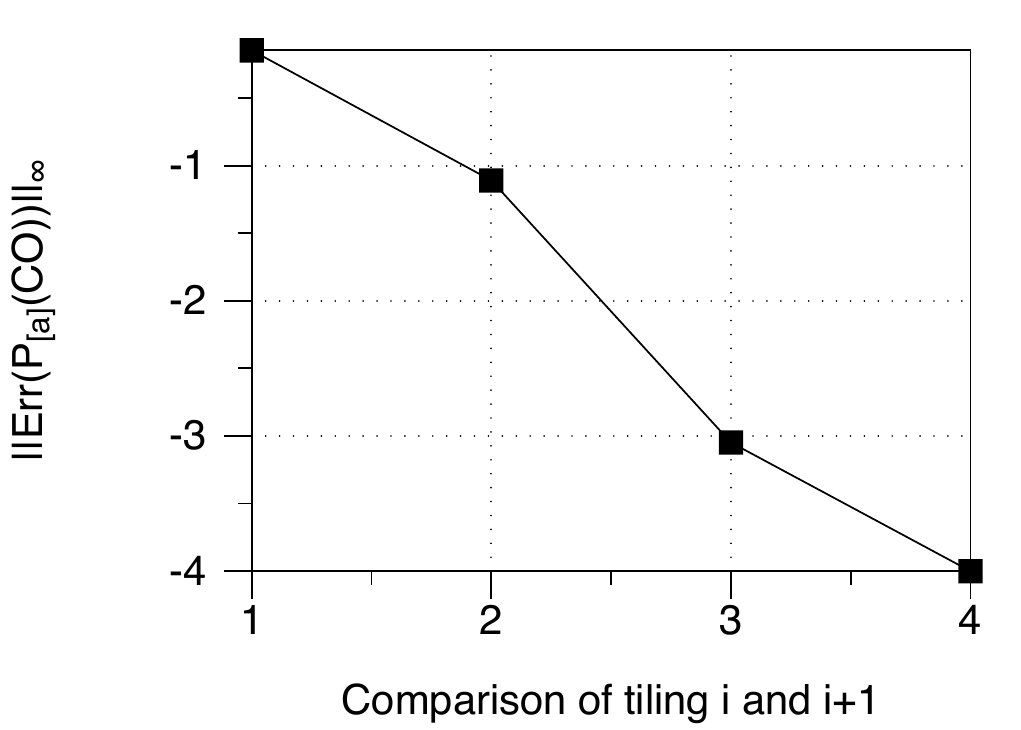}(a)}
\mbox{\includegraphics[width=5cm]{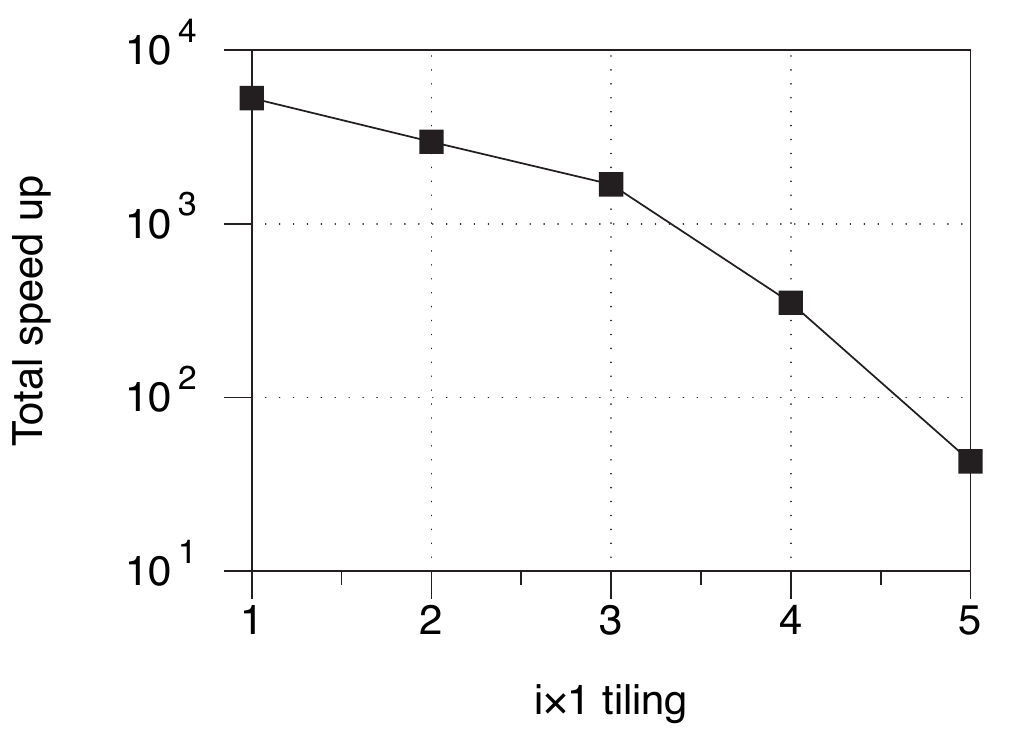}(b)}
\caption{For the 1D system, { we take the difference between the steady state values of $P($CO$)$ over successive tilings in the hierarchy, and demonstrate how the error between tilings deteriorates as the hierarchy grows} (a).  The total speed{-up} between an $i\times1$ tiling and the KMC simulations over 10 different parameter runs is shown (b).  The $5\times1$ tiling runs 43 times faster than the KMC simulations.}
\label{fig:errandsp1D}
\end{figure}

\begin{figure}
\includegraphics[width=8cm]{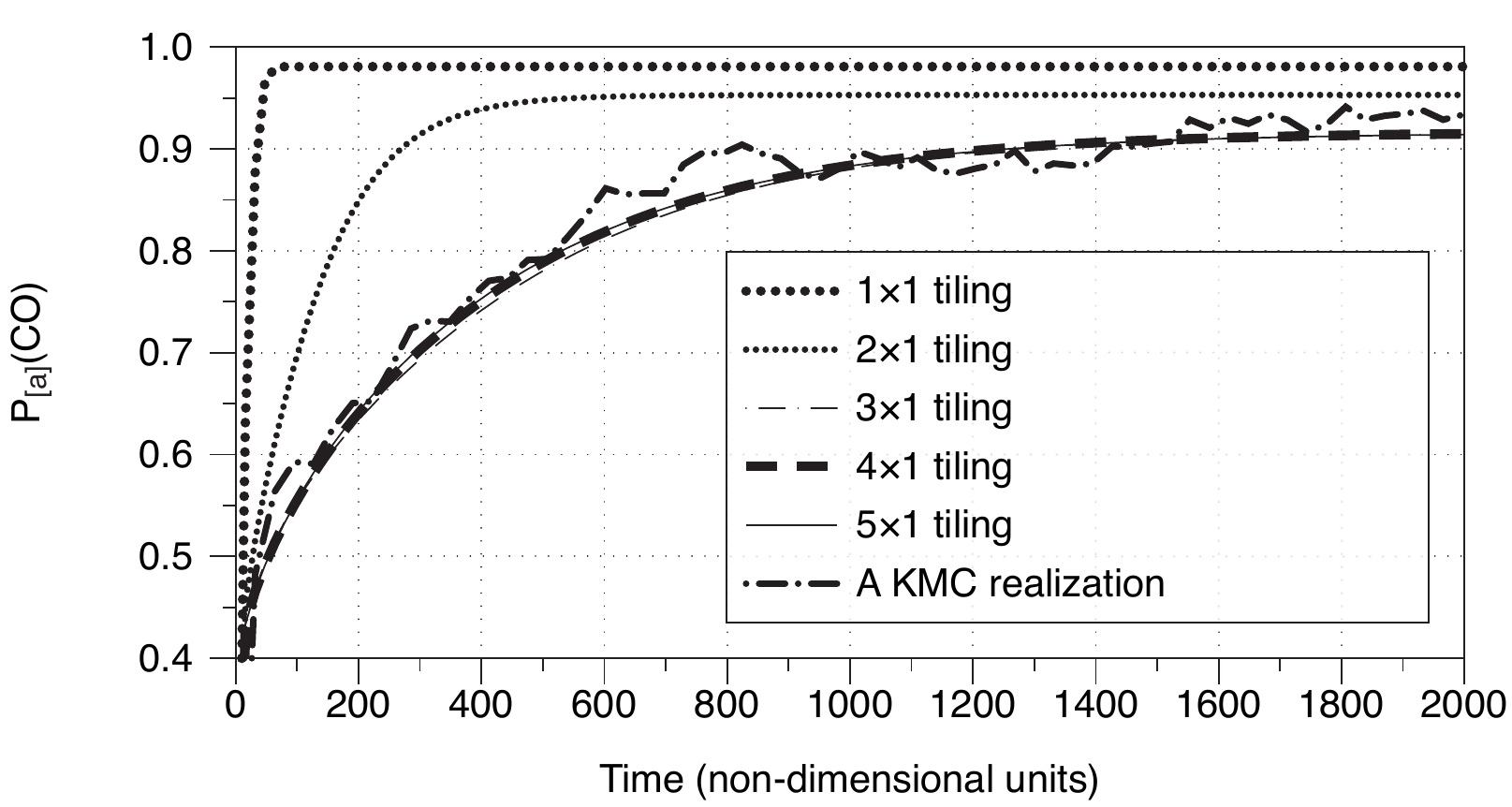}
\caption{In the 1D system, dynamics of all tiling types are compared to a single KMC realization for $k_{\emptyset\rightarrow CO} = 10^{-0.2}$.}
\label{fig:1Ddyn}
\end{figure}

With the results from the ODEs, we use the information on the larger tiles to determine the probability of finding a CO on any given site, and plot the steady state of this value for each {level in the hierarchy} in Figure \ref{fig:ss1d}.
Steady state values for the KMC simulations are found by approximating the time scale of system equilibration, $T$, predicted by the higher dimensional tilings, and then running the {set of equations} for $10T$ and averaging the results from $t\in[5T,10T]$.
For $(4\times1)$ and larger tilings, the approximated master equation dynamics fall less than a quarter of the standard deviations from the means of the KMC simulations.
The $(3\times1)$ tiling performs very well lying within two standard deviations of the KMC statistics over all tested parameter values.
We next plot the infinity norm of the absolute error over all predicted steady states between successive tilings, and we find that the developed method converges with order $1.35$ (Figure \ref{fig:errandsp1D}).
Although we have statistically determined the steady states from the KMC simulations, we have not performed a statistical sampling for the KMC time dynamics in this test.  We note, however, that the relaxation time scales appear to match precisely for the higher dimensional tiles ($n>3$) and the KMC simulations; a single KMC realization is compared to all of the tilings in Figure \ref{fig:1Ddyn}, for $k_{\emptyset\rightarrow\text{CO}}=10^{-0.2}$.  As expected, the execution times of the ODE's are far faster than KMC simulation (see Figure \ref{fig:errandsp1D}).  In the case of the $(3\times1$) tiling, the approximated {equations} run 1690 times faster than the KMC simulations, and in the more accurate case of the $(4\times1$) tiling, the approximated {equations} run 352 times faster than the KMC simulations.  We note that the tiling associated ODEs are solved using serial CPU execution, whereas the KMC simulations exploit parallel capabilities of GPUs, and thus the true acceleration of our methods can be even greater than what we have presented (exact performance figures depend to a large extent on computer architecture; all tests were run on a standard early 2013 15" MacBook Pro).

\subsubsection{Identical cus and bridge sites}\label{sec:IDbc}
\begin{figure}
\mbox{\includegraphics[width=8cm]{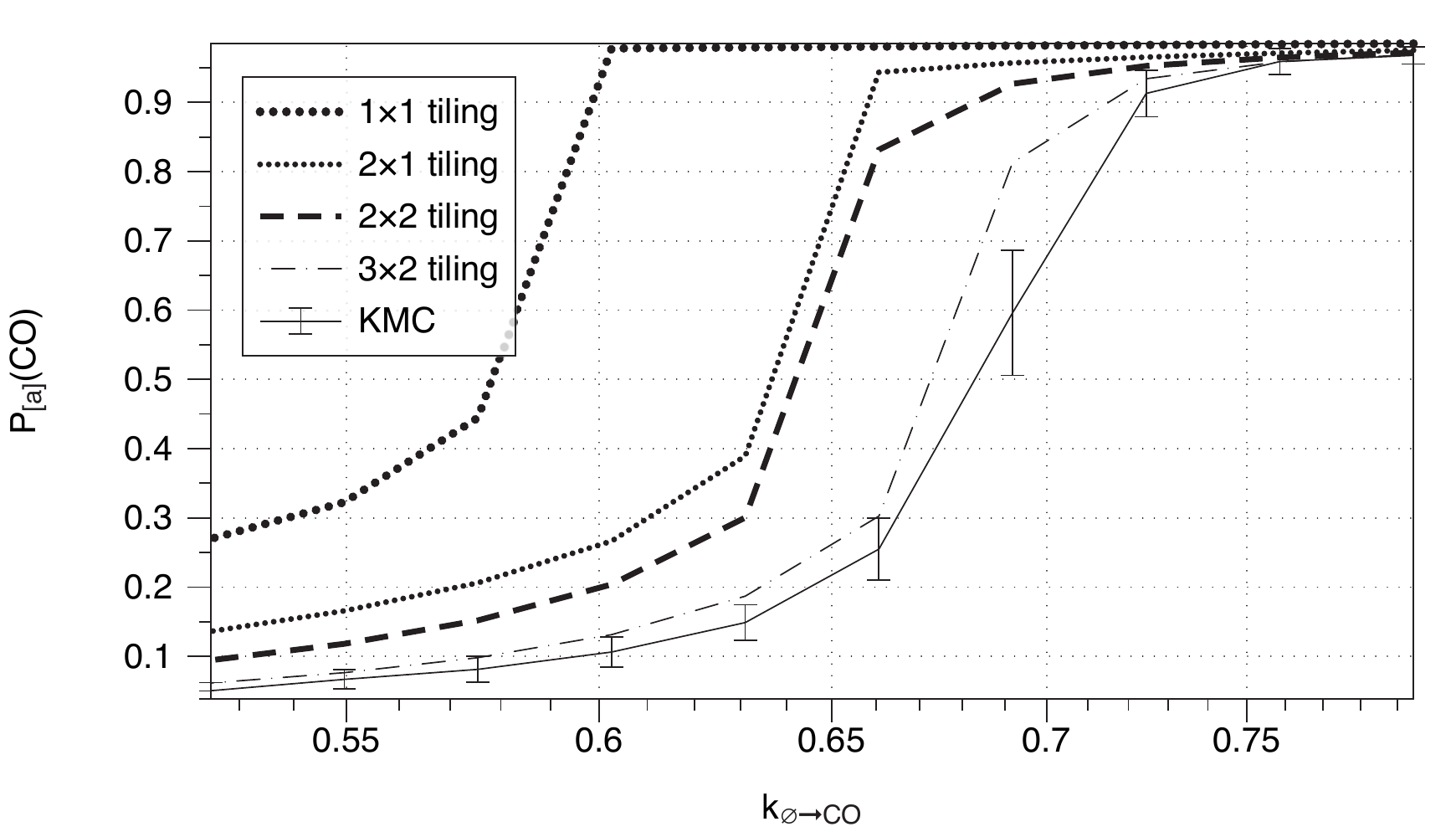}(a)}
\mbox{\includegraphics[width=4.8cm]{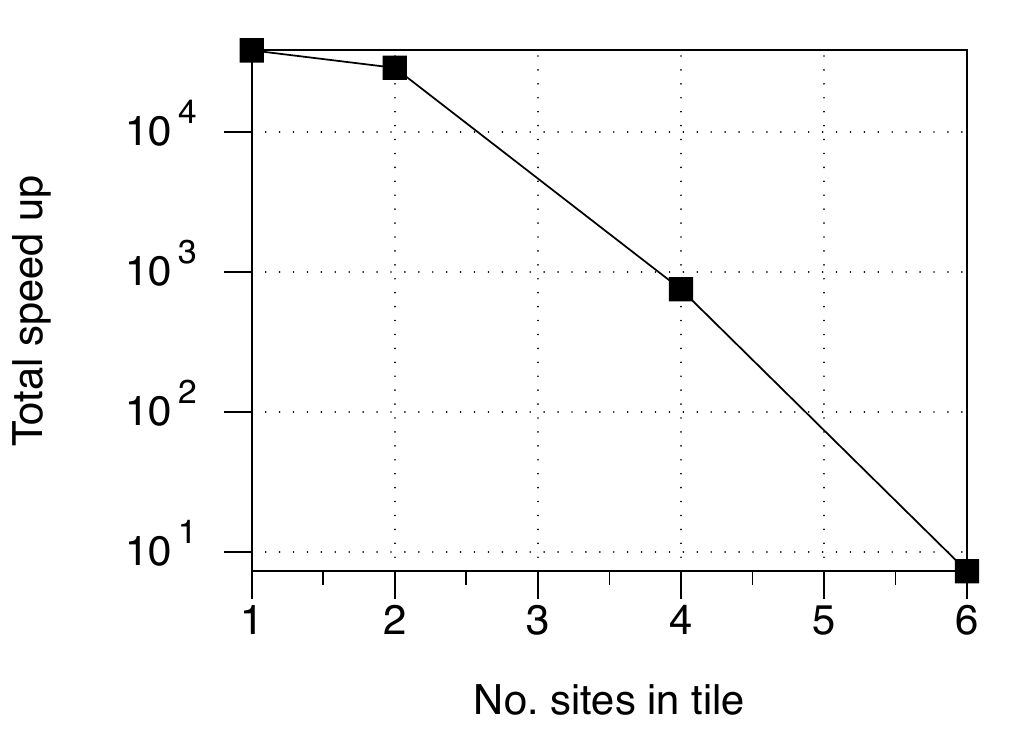}(b)}
\caption{Steady states (a) and relative speed up (b) in the two-dimensional case in which bridge and cus sites are treated identically.  The speed up is plotted based on the number of sites per tile.  The $(3\times2)$ (6 tile sites) approximation has the slowest speed up and runs 7.7 times faster than the KMC simulation over all 10 data runs.}
\label{fig:2DSSandspu}
\end{figure}

Next, we analyze the two dimensional system in which bridge sites are treated identically to cus sites.  In this case we run the KMC simulations for a $32\times32$ periodic grid, and compare $(1\times1)$, $(2\times1)$, $(2\times2)$ and $(3\times2)$ tilings which yield $3,3^2,3^4$ and $3^6$ dimensional {equation}s respectively.  We again compare the steady states of the tiling approximations with results from KMC simulations and find that only the $(3\times2)$ tiling approximation lies within two times the standard deviation of the KMC results for all parameters (see {\color{\red}\sns{fig} Figure} \ref{fig:2DSSandspu}).  The $(2\times2)$ tiling does not show significant improvement over the $(2\times1)$ tiling.  Additionally we plot the speed up in Figure \ref{fig:2DSSandspu}.  The $(3\times2)$ tiling approximation runs 7.7 times faster than the KMC simulation.

\begin{figure}
\includegraphics[width=8cm]{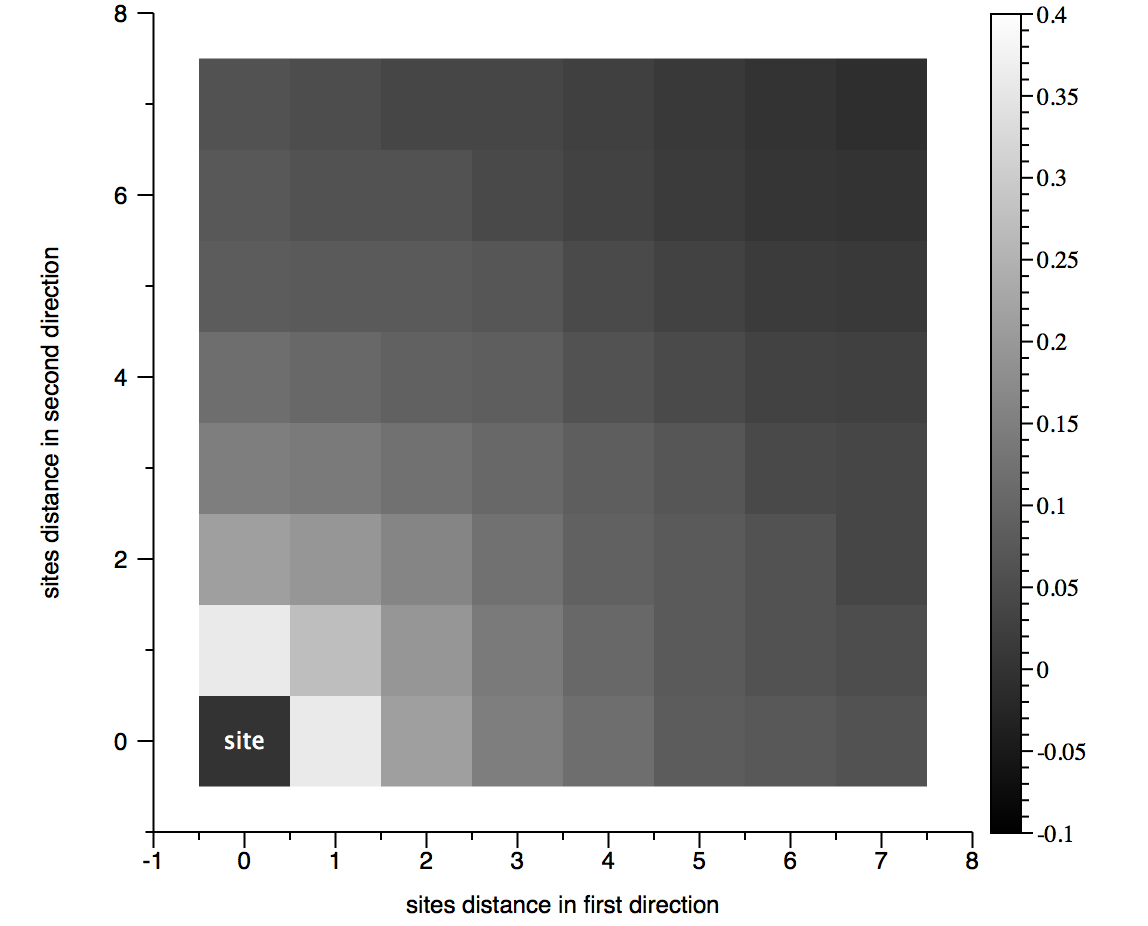}
\caption{ We plot the correlation of states between neighbors located a distance of $i$ lattice sites to the right and $j$ lattice sites above.  The correlations are taken at steady state and $k_{\emptyset\rightarrow CO} = 10^{-0.16}=0.69$.  We note that correlation decays slowly as a function of distance implying that we must take a large element in the hierarchy in order to predict accurate system dynamics.}
\label{fig:cor2D}
\end{figure}

We do not test larger tilings as the {ODE} matrices quickly become too large for the LSODA method to approximate the Jacobian.  We determine the spatial correlations where the approximations have maximal error ($k_{\emptyset\rightarrow CO} = 10^{-0.16}=0.69$; see Figure \ref{fig:cor2D}).  We find that at steady state, the spatial correlations die down over four nearest neighbors, and determine that the fifth nearest neighbors have correlations that are less than 10\% of nearest neighbor correlations.  Thus we conjecture that $4\times4$ or $4\times3$ tilings to be within the mean of the KMC simulations.  We do not, however, test this conjecture as the {number of equations is} too large for the memory requirements of the LSODA routine in constructing the approximated Jacobian which would require $3^{16}\times3^{16}$ and $3^{12}\times3^{12}$ dimensional arrays respectively.  {\color{\red} Below in the current section and in the discussion, we suggest several methodologies of reducing the number of dimensions for these larger systems, however do not formally investigate these methods in the current work.}

\subsection{Catalytic oxidation of CO by a RuO$_2$(110) surface}\label{sec:RuO2}
We next test the tiling approximations for the catalysis problem described in refs. \onlinecite{ReuterScheffler:06,Temeletal:07}. 
In this system $\mathcal{S} = \{\emptyset,$ O, CO$\}$, $\mathcal{T}=\{$bridge,cus$\}$.
{ Although previous gPK methods claim to be able to handle different site types, to our knowledge there has been no work that has examined these models on a regular lattice made up of different site types (however there have been gPK models on lattices with randomly active/inactive site{\color{\red}s} \cite{Mai:93b} as well as {\color{\red} disordered} heterogenous lattices \cite{Cortes:01}).  In the current tiling framework, however, such an extension becomes natural.} Using the formalism above, we compare $(1\times1)$ and $(2\times1)$ tiling approximations with results from KMC simulations.
The tiling types $\Xi$ may again be mapped directly to the site types which are $\mathcal{T}'=\{$[b],[c]$\}$ for a $(1\times1)$ tiling and $\mathcal{T}'=\{$[b,b],[b,c],[c,c]$\}$ for a $(2\times1)$ tiling.
These approximations result in { a} $2\times3$ and $3\times3^2$ dimensional {ODE}.
We repeat one of the numerical experiments from ref. \onlinecite{Temeletal:07}, in which we assume that the partial pressure of CO$_2$ is zero ($p_{\text{CO}_2}=0$), fix the partial pressure of O$_2$ to be 1atm ($p_{\text{O}_2}=1$atm) and determine the system evolution for a variety of partial pressures of CO ($p_\text{CO}$), ranging from 0.5 to 50 atm (21 partial pressures evenly partitioned on a log scale).  The temperature of the system is taken to be 600K.
Reaction rates are taken from ref. \onlinecite{Temeletal:07} {; for self consistency we report these values in table \ref{tab:param}}.

\begin{table}
{\caption{\label{tab:param}We present the parameters used in simulating the oxidation of CO on RuO$_2$.  The table is a partially reconstructed table from table 1 found ref. \onlinecite{Temeletal:07}.  Parameters that differ by more than 16 orders of magnitude (machine $\epsilon$) from the largest parameter values are set to zero.}}
{
\begin{tabular}{l | l}
\hline\hline 
Process & Rate (s$^{-1}$)\\
\hline
CO$_g$+$\emptyset_b \rightarrow$ CO$_b$ & $7.2\times10^8\times \frac{p_{\text{CO}}}{7}$\\
CO$_g$+$\emptyset_c \rightarrow$ CO$_c$ & $7.2\times10^8\times \frac{p_{\text{CO}}}{7}$\\
O$_{2}$$_g$+$2\emptyset_b \rightarrow$ 2O$_b$ & $9.7\times10^7$\\
O$_{2}$$_g$+$2\emptyset_c \rightarrow$ 2O$_c$ & $9.7\times10^7$\\
O$_{2}$$_g$+$\emptyset_b+ \emptyset_c \rightarrow$ O$_b$+O$_c$ & $9.7\times10^7$\\
CO$_b \rightarrow$ CO$_g$+$\emptyset_b$ & $2.8\times10^4$\\
CO$_c \rightarrow$ CO$_g$+$\emptyset_c$ & $9.2\times10^6$\\
2O$_b \rightarrow$ O$_{2}$$_g$+$2\emptyset_b$ & $0$\\
2O$_c \rightarrow$ O$_{2}$$_g$+$2\emptyset_c$ & $2.8\times10^{1}$\\
O$_b$ + O$_c \rightarrow$ O$_{2}$$_g$+$\emptyset_b+\emptyset_c$ & $0$\\
CO$_b$ + O$_c \rightarrow$ CO$_{2}$$_g$+$\emptyset_b+\emptyset_c$ & $1.2\times10^6$\\
CO$_b$ + O$_b \rightarrow$ CO$_{2}$$_g$+$2\emptyset_b$ & $1.6$\\
CO$_c$ + O$_c \rightarrow$ CO$_{2}$$_g$+$2\emptyset_c$ & $1.7\times10^5$\\
CO$_c$ + O$_b \rightarrow$ CO$_{2}$$_g$+$\emptyset_b+\emptyset_c$ & $5.2\times10^2$\\
\hline\hline
\end{tabular}}
\end{table}

\begin{figure}
\begin{center}
\mbox{\includegraphics[width=8cm]{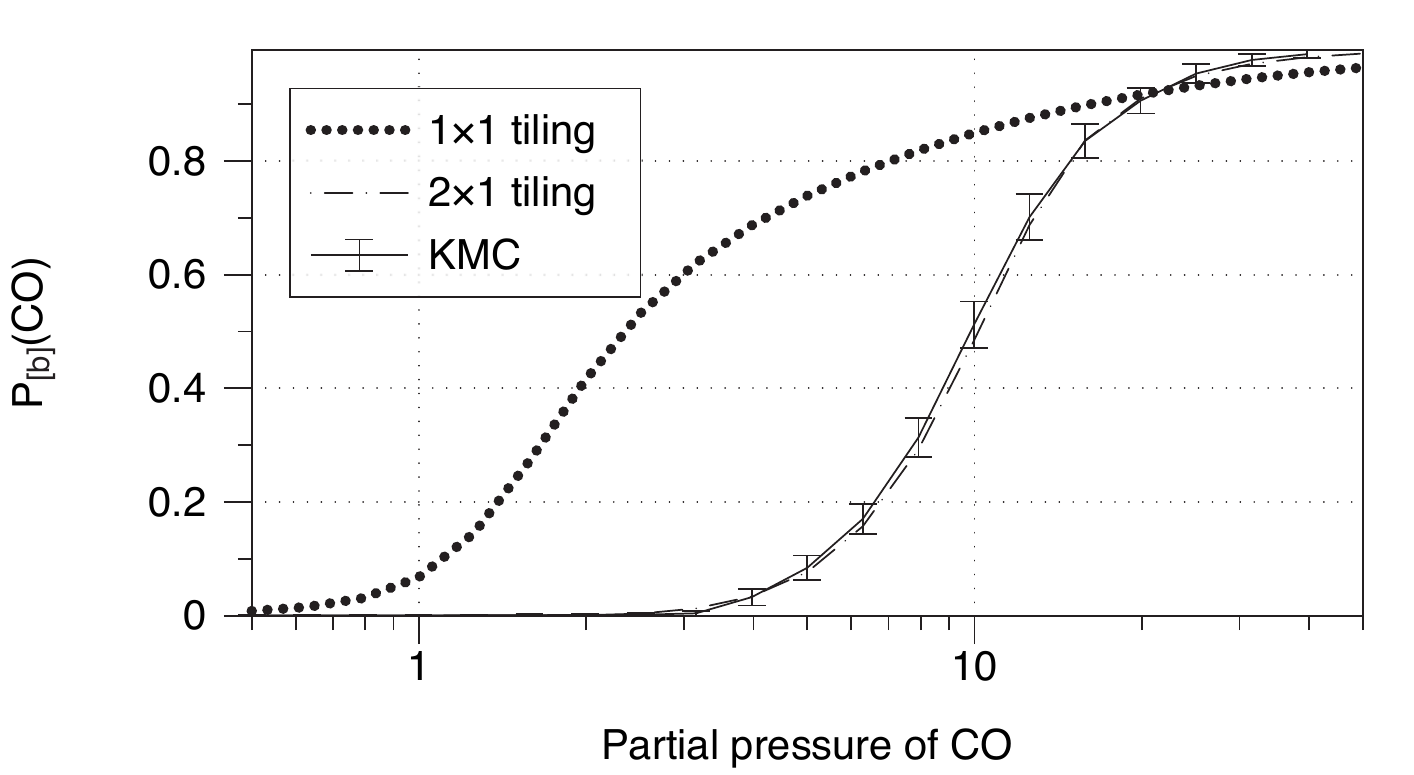}(a)}\\
\mbox{\includegraphics[width=8cm]{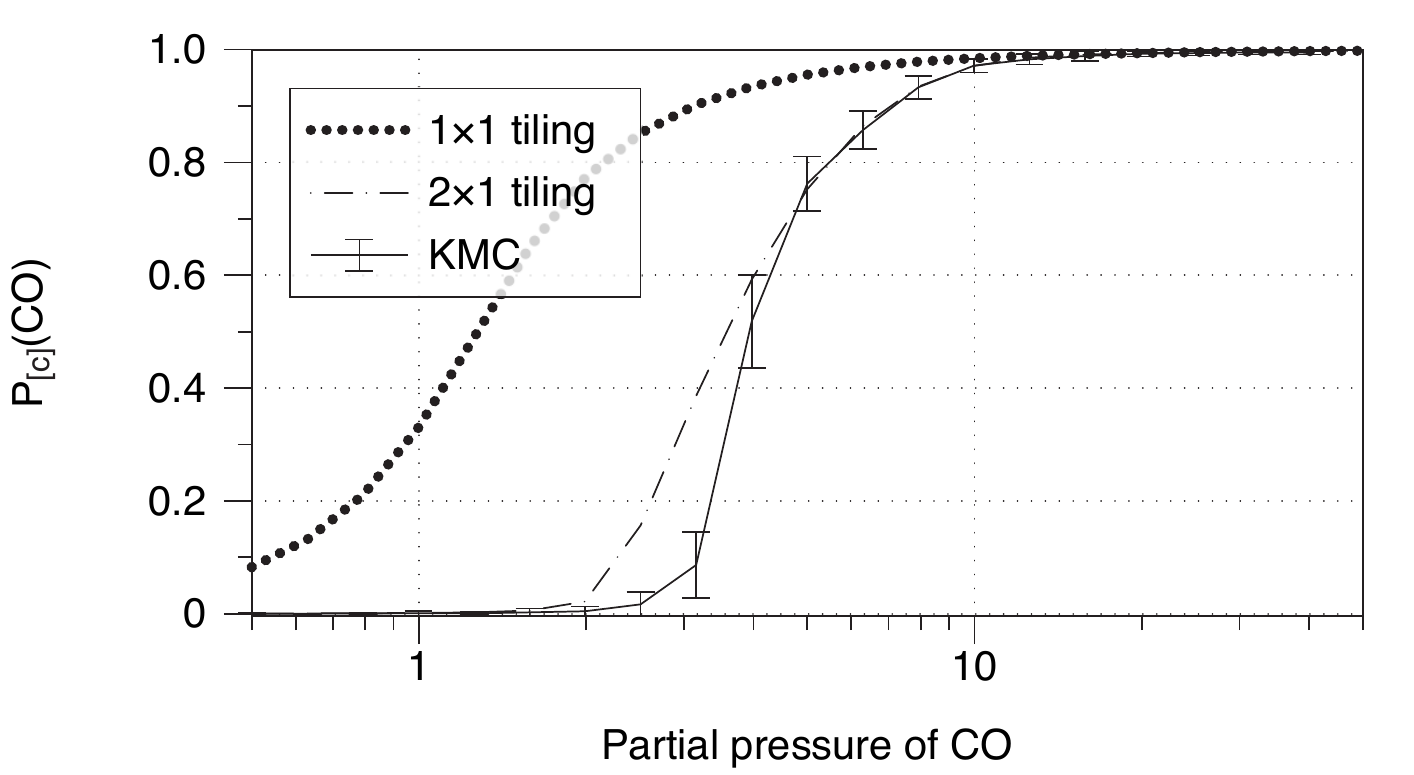}(b)}\\
\caption{The steady state probability of occupancy by CO is shown for bridge sites (a) and cus sites (b).  Notice that the $(2\times1)$ tiling approximation equation fits with in the error bars of the KMC simulation for the bridge sites, but does not for some regions of the cus sites.  The traditional $(1\times1)$ tiling predicts values and behavior with large error.}
\label{fig:brsites}
\end{center}
\end{figure}

To determine the accuracy of the $(1\times1)$ and $(2\times1)$ tilings, KMC simulations are performed on a $60\times60$ grid and 98 runs are completed at each partial pressure.
On the bridge sites the $(2\times1)$ tiling approximation falls within a standard deviation of the mean KMC results for site occupations (see Figure \ref{fig:brsites}).
On cus sites the $(2\times1)$ tiling approximation fails for partial pressures greater than 2 and less than 5 atm (see Figure \ref{fig:brsites}).
The $(2\times1)$ tiling approximation, however, demonstrates a vast improvement from the $(1\times1)$ tiling approximation (PK) model; far more so than the parameters of the previous section.  A similar study was performed in ref. \onlinecite{Temeletal:07} in which the authors compared the PK model with KMC simulation and their results match ours.  {We also note that there is existing evidence that the pair model will perform well in this scenario from ref. \onlinecite{Matera:11}, in which the authors demonstrate that by better approximating the pair probabilities from the probability distribution on cus sites for low values of $p_{co}$, a modified PK model can perform very well in terms of predicting turnover efficiency up to the point where O and CO begin to coexist on the surface. }

\begin{figure}
\begin{center}
\includegraphics[width
=8cm]{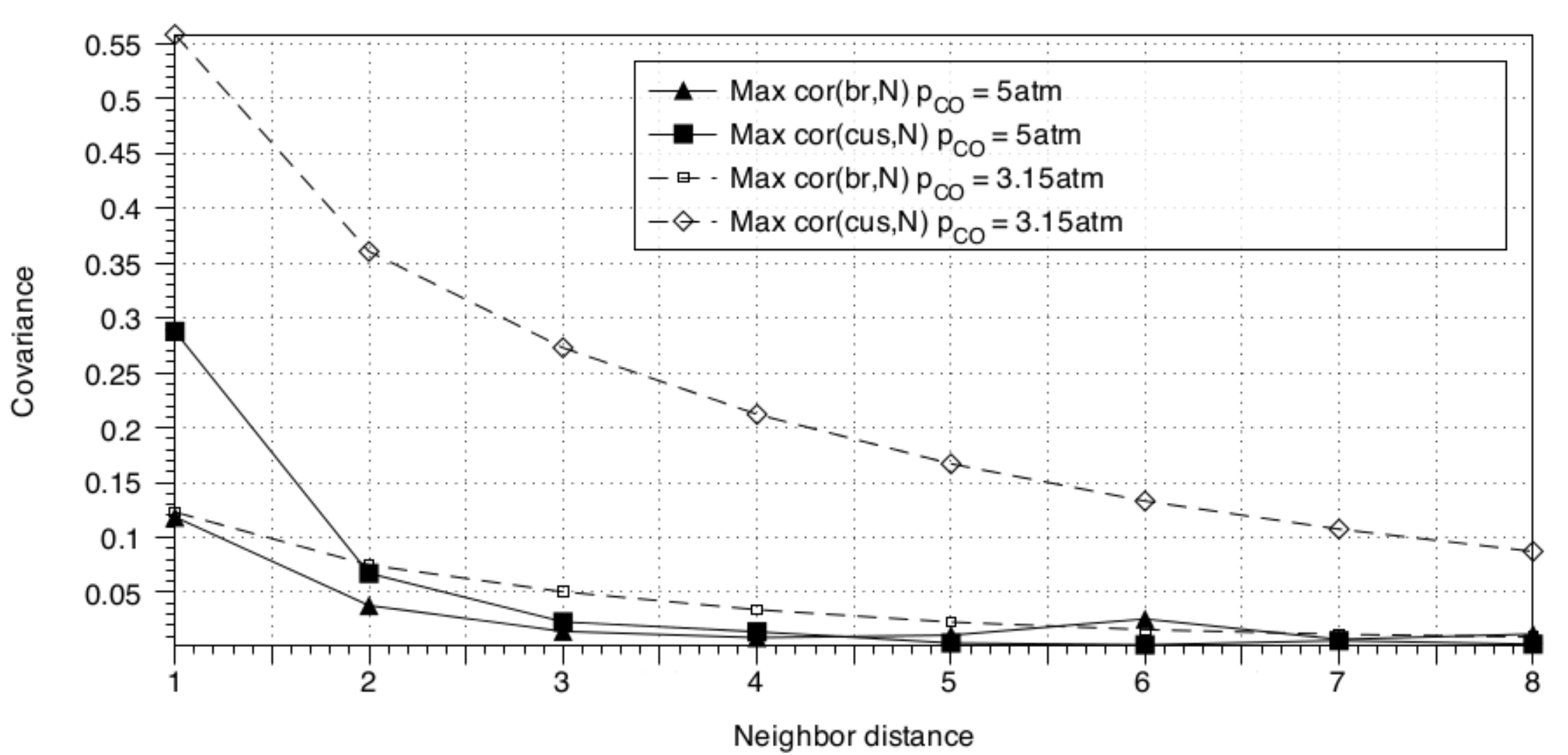}
\caption{The maximum covariance as a function of distance from a site is given for $p_{\text{CO}}=5$atm and $3.15$atm.  The former value shows low correlations beyond nearest neighbors which explains why the $2\times1$ tiling scheme accurately predicts the system dynamics at this partial pressure.  The latter value demonstrates high correlations even up to eight neighbors away { which explains why the $2\times1$ tiling scheme was insufficient in this regime}.}
\label{fig:covar}
\end{center}
\end{figure}

\begin{figure}
\begin{center}
\includegraphics[width=8cm]{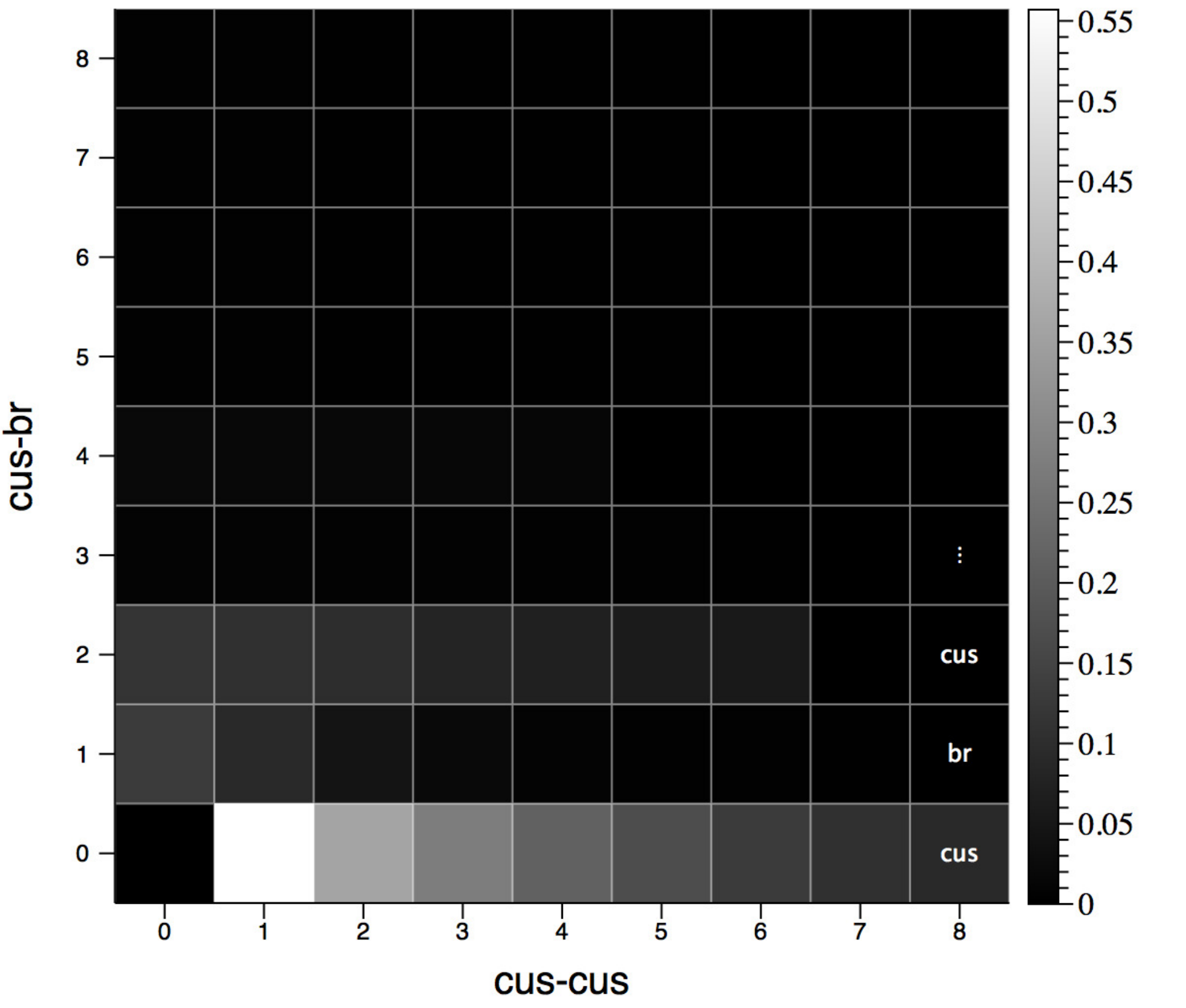}
\caption{The correlation for a cus site is decomposed based on the system geometry.  The horizontal axis shows a line of constant cus sites and the high correlation that persists in this line.  In the vertical direction, the first bridge and cus site away show non-negligible correlation to the cus site at the origin.}
\label{fig:cuscor}
\end{center}
\end{figure}

\begin{figure}
\begin{center}
{\includegraphics[width=8cm]{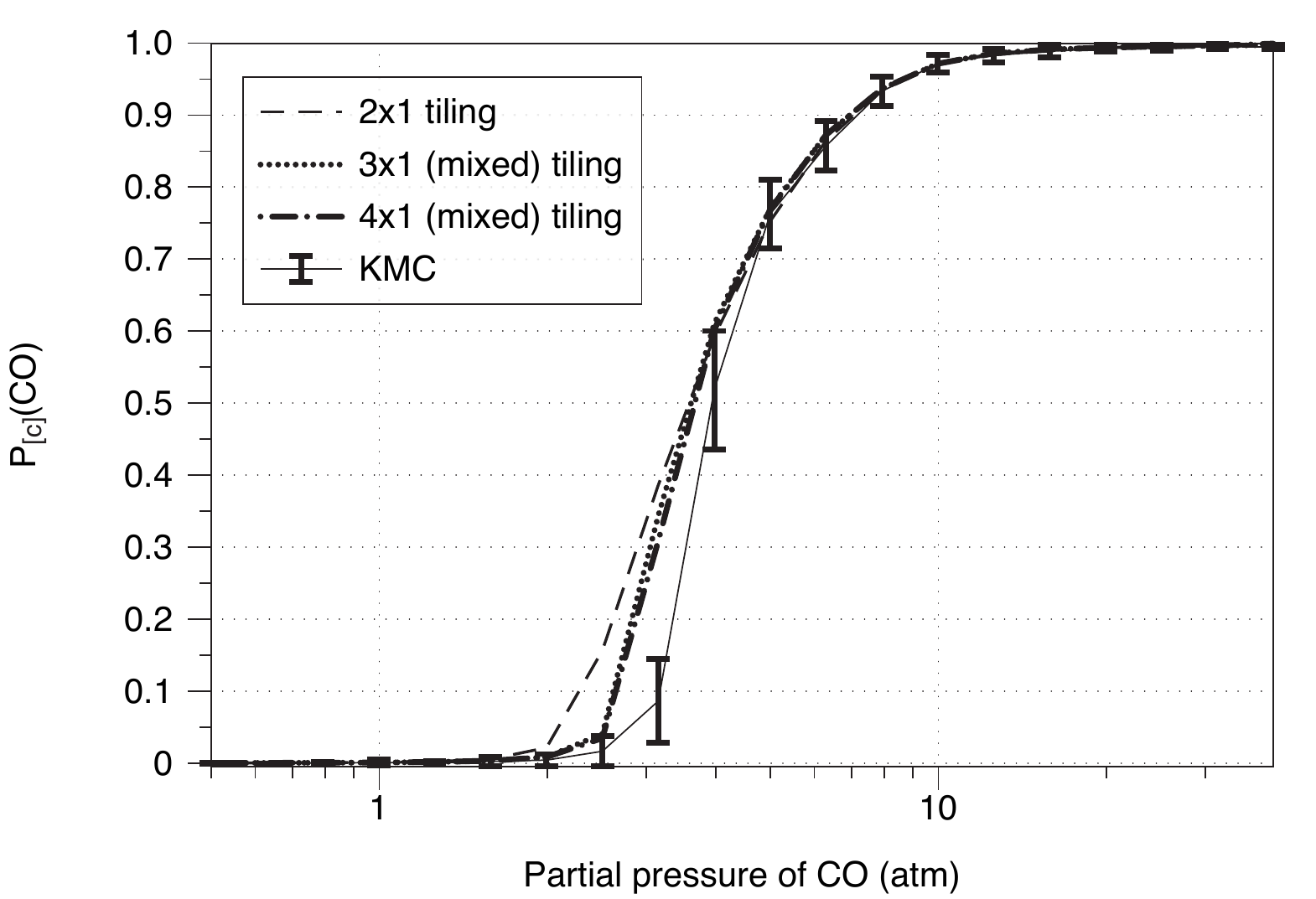}}
\caption{Adding a mixed tiling scheme that accounts only for the spatial correlations in the cus-cus direction shows an improvement from the $(2\times1)$ tiling scheme, however we do not see convergence when the partial pressure of CO is 3.15.}
\label{fig:mixedtile}
\end{center}
\end{figure}

To determine the approximated size of a tiling that would lead to an accurate description of the system dynamics, we again examine the length scale correlations at steady state as a function of partial pressure.
Within the KMC simulations at a partial pressure of $p_{\text{CO}}=3.15$atm, we find that the bridge-bridge correlations die off nearly completely after the nearest neighbor; the cus-bridge pairs, however, are significantly correlated up to two neighbors away, while the cus-cus pairs are significantly correlated beyond 8 neighbors away (see Figures \ref{fig:covar} and \ref{fig:cuscor}).
This data supports the observations that the predicted dynamics for bridge sites is accurate for a $(2\times1)$ tiling, whereas the predicted dynamics for the cus sites is not.
To accurately capture the system dynamics at this partial pressure, we would need either an unmixed $9\times2$ tiling approximation which would result, at minimum, in a $|\mathcal{S}|^{9\times2}=3.87\times10^{8}$ dimensional {ODE}.  We could also potent{i}ally use a mixed $(9\times1)$ (and $(1\times9)$) system which would  lead to a $|\Xi|\times|\mathcal{S}|^{9} \approx 6\times10^4$ dimensional {ODE} which is far more tractable.  Finally it is also possible to use a more complex mixed tiling system such that we use $(2\times1)$ tiles for bridge-bridge and bridge-cus connections and $(n\times1$) tiles in the cus-cus direction, which would lead to a $2\times |\mathcal{S}|^{2}+|\mathcal{S}|^{n}$ dimensional {ODE}; for $n=9$, this gives a $\sim2\times10^4$ dimensional {ODE}, which is far more {tractable} still.  The first method leads to a {set of equations} which is intractably large.  The second {corresponding ODE} is numerically tractable and the typical system we have been using in the present work.  The third system is yet a new tiling structure which we begin to explore {\color{\red} by increasing the size of an $(n\times 1)$ tiling in the cus-cus direction until we observe the hierarchy to converge.  We note that when $n=2$, the mixed tiling scheme is equivalent to the $2\times1$ scheme examined above. \sns{ with $n=\{2,3,4,5\}$, the $n=2$ case being equivalent to the $2\times1$ scheme that we have examined above.}} We check for consistency over all single site projections of the probability of finding CO on cus sites for each case and verify that the consistency criteria is satisfied in these test cases.  We plot the results found at steady state in figure \ref{fig:mixedtile}.  We find an improvement in the mixed tiling scheme, {\color{\red} and note that the mixed tiling scheme has converged when $n=4$ (found by comparing with the $n=5$ case).  We note} however {\color{\red}\sns{note}}that we do not see {\color{\red} convergence that accurately captures the predictions of KMC simulation. \sns{  full convergence having only gone up to a $(5\times1)$ tiling scheme in the cus-cus direction.}}   As in the previous section, we compare the speed up of the generalized PK models with a single KMC run and determine the average speed up over all examined partial pressures.  We note that we have taken 98 KMC simulations and thus the actual average speed up in our computations is 98 times greater than what is presented.  We display the speed up in figure \ref{fig:mixedtilespeedup} and find that the $(2\times1)$ tiling scheme runs 4500 times faster than a single KMC realization.  The $(5\times1)$ mixed tiling scheme yields a speed up of 4.5 for a single KMC simulation run.

\begin{figure}
\begin{center}
{\includegraphics[width=8cm]{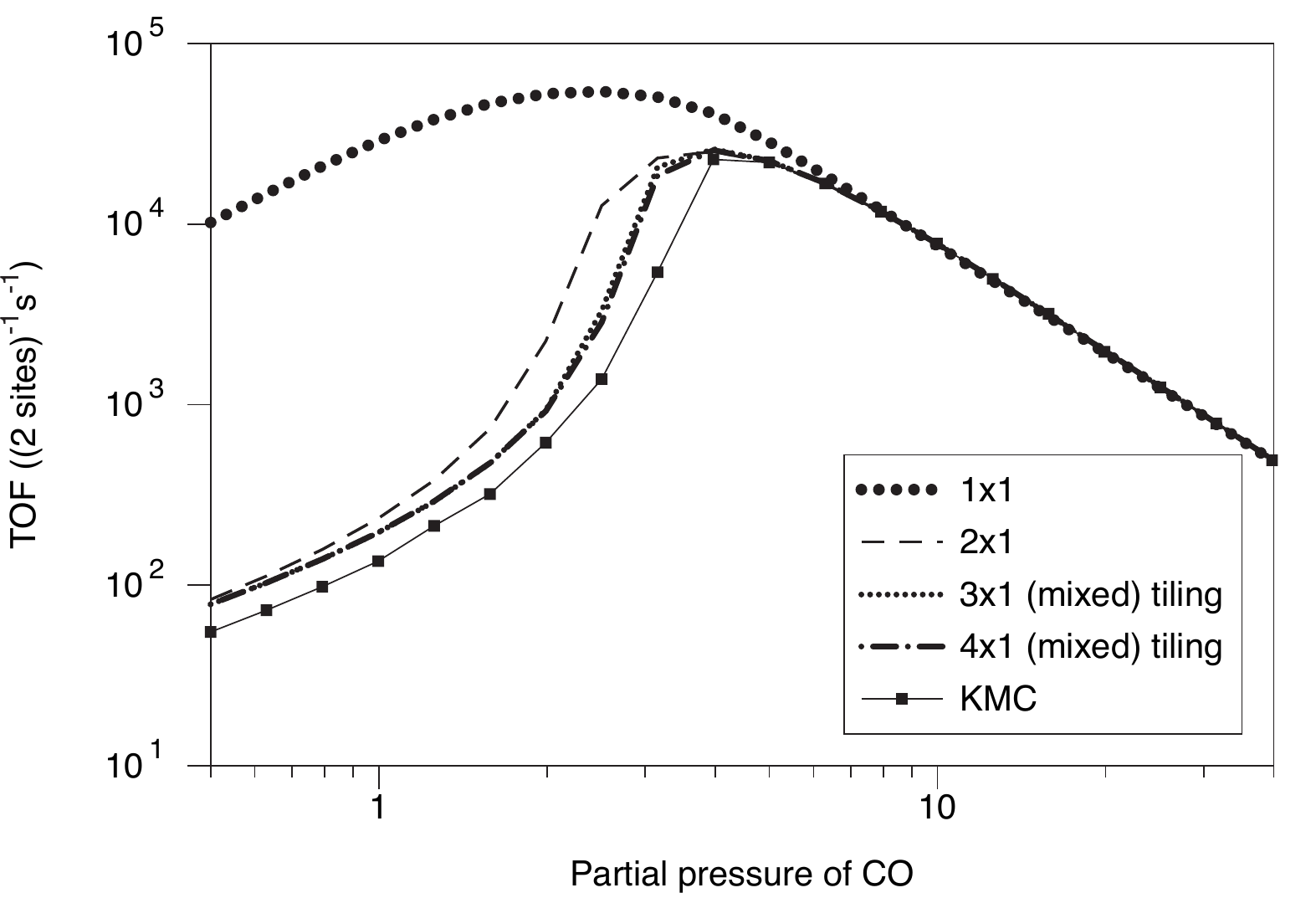}}
\caption{The TOF is presented as a function of the partial pressure of CO.  See text for details.}
\label{fig:TOF}
\end{center}
\end{figure}

\begin{figure}
\begin{center}
\includegraphics[width=4.8cm]{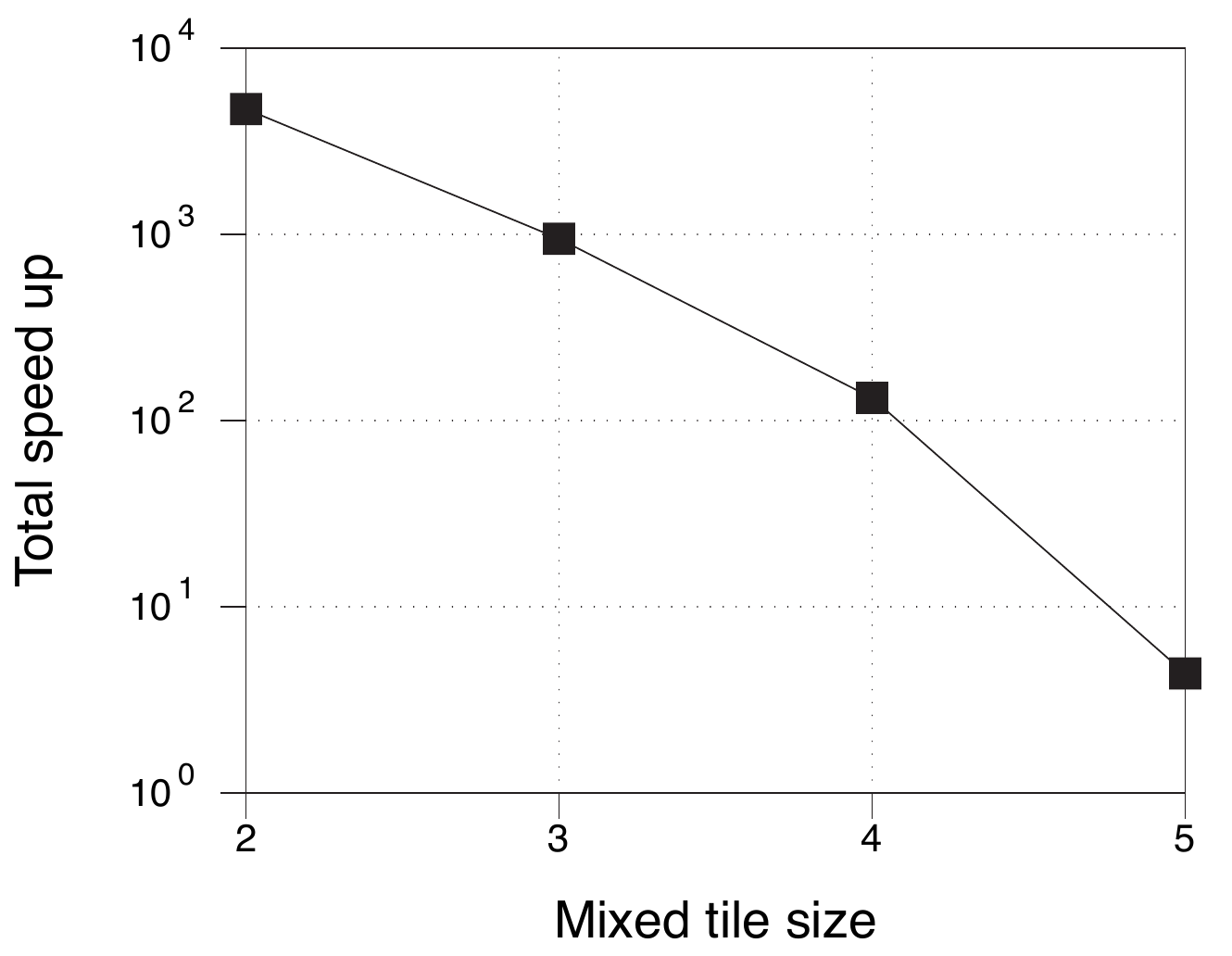}
\caption{We find significant computational speed up when comparing the mixed tiling schemes with KMC computational times.  We find a speed {up} of 4.5 for the largest case when mixing $(5\times1)$ tilings.}
\label{fig:mixedtilespeedup}
\end{center}
\end{figure}

{ Finally, we examine the accuracy of the predicted turnover efficiency.  In applications the true quantity of interest is the rate at which CO is oxidized in to CO$_2$.  This quantity is called the turnover efficiency (TOF) which is a measure of how often CO is oxidized on the surface.  Here we define the TOF as the number of reactions per two sites per second and write it as

\begin{align}
{\color{\red}TOF} & {\color{\red}\equiv\sum_{i,j\in\{br,cus\}}  k_{\xi_{ij}(CO,O)\rightarrow\xi_{ij}(\emptyset,\emptyset)} p_{CO\: O}^{ij}}
\end{align} 
{\color{\red}where we have used the notations defined in section \ref{subsec:2x1}}. We plot and compare the TOF for the KMC simulations, the PK model, the $2\times1$ tiling and the mixed $3\times1$ and $4\times1$ tilings in figure \ref{fig:TOF}.  We find that the prediction for the location and value of the peak TOF is significantly improved even in the $2\times1$ tiling scheme.  We note further that for the oxygen poisoned regimes for small partial pressures of CO, the tiling model performs very well, and is comparable to that of the problem specific modified PK model presented in ref. \onlinecite{Matera:11}.  The slight over prediction is consistent with the single site predictions, as CO is over estimated on the cus sites (see {\color{\red}\sns{f}F}igure \ref{fig:mixedtile}).  The over prediction is thus due to the abundance of oxygen and the relatively large rate $k_{CO_{cus}O_{cus}\rightarrow \emptyset_{cus}\emptyset_{cus}}$.
}

With the two tiling systems that we have presented, we have confirmed that the hierarchy of {kinetic equations} leads to improved predictions for realistic surface dynamics and have shown that this occurs even with small improvements within the hierarchy.  We have also introduced the idea of mixed tiling systems and shown how they may be used to introduce improvements in the accuracy of the dynamics.

%
%

%

\section{Discussion}
We have developed a method of approximating the master equation for systems that are assumed to be translationally invariant with finite spatial correlations.  In principle, this {hierarchy} will always converge to the master equation and we have shown this in one- and two-dimensional examples.
The computational cost of the developed method has been explored and {compared with KMC simulation} on the test examples and we find that for smaller {elements of the hierarchy}, we see significant reductions in computational time.
We have also shown that the methods quickly lead improvements on a realistic example of oxidation of CO on RuO$_2$(110), and have observed that a modest improvement to $(2\times1)$ tilings captures many of the important system dynamics within the parameters considered.  { Although the $(2\times1$) tiling scheme is similar to the gPK model presented in ref. \onlinecite{Mai:93}, the present work, to our knowledge, is the first to explore the surface reaction dynamics of gPK {equations} within the context of a non-uniform or non-randomized lattice.  Furthermore, we have introduced the concept of a mixed tiling, and have shown that mixed tiling schemes can lead to a more accurate description of the dynamics of the master equation.  
}

The ability to extend models to larger tilings provides a means to hypothesis test PK models on smaller tiles{, as well hypothesis test gPK models by examine larger tilings}.
In any pursuit in which one hypothesizes that a  gPK model provides a suitable model, the current methodology provides a fast method to justify this choice of model by testing this hypothesis with extended tiling systems.  Should the dynamics change significantly between the smaller and larger tilings, we can reject the method.  Although this provides a sufficient tool for hypothesis rejection, it is an open and interesting question to ask that if we do not see improvement between a smaller and larger tiling system, does this mean that the method has converged or are there local plateaus (i.e. is the condition necessary)? { For example, although we did not see convergence in the mixed tiling scheme on the realistic example, we did see progressively more accurate schemes when both spatial directions were accounted for in the tiling scheme.  A rigorous framework describing the situations in which the tiling schemes will converge to the appropriate dynamics will be in an important next step in developing this work.}

This hierarchy of gPK models may also be used to fit parameters from observed experimental data.  These parameter estimates can be similarly tested by examining larger tilings.  If the parameters change, we can conclude that longer range spatial correlations play a significant role in the surface dynamics, however if the parameter estimates do not change it is still an open question as to whether or not we can conclude these are accurate surface parameters.  If this question can be answered in the affirmative, we can then use transition state theory (TST) to predict energy barriers and energy differences between bound and unbound site states, and also predict transition rates over all temperatures.
The two open questions presented in this and the above paragraph will be the subject of a future investigation.

{ We note that it is possible for the gPK model to be too computationally expensive in order to select a corresponding tile that will guarantee convergence as is demonstrated in the 2D uniform surface example of section \ref{sec:IDbc}.  To potentially resolve this issue we have noted the possibility of introducing mixed tiling surfaces and have demonstrated the possibility for improved accuracy.  We propose the idea that a mixed tiling search algorithm may be able to determine appropriate directions on which to increase a mixed tiling scheme, but save such a development for future work.  We also note that speed-ups beyond what we have presented in the present work are possible by (1) solving resulting ODEs with a quasi-Newton method rather than via direct integration of the ODEs from some initial condition, (2) by utilizing symbolic programing to reduce the number of degrees of freedom via the consistency restraints found in section \ref{sec:constraint}.  We save both of these tasks for future work, but note that we have begun to develop a code base for this algorithm and an example of the code is available at https://github.com/gjherschlag/mixedtiling\_eg.  These speedups and savings in memory will allow us to search larger elements of the hierarchy that may allow us to reach convergence for a wide class of problems with large computational savings.}

The methodology here has been tested in the context of constant rate coefficients so that Equations \ref{eqn:inmatrxgenreduct} and \ref{eqn:outmatrxgenreduct} may be simplified to Equations \ref{eqn:inmatrxreduct} and \ref{eqn:outmatrxreduct}.  In many interesting catalysis reactions, rate reactions will change based on local spatial correlations.
Although we have not investigated such mechanisms in the current work, it will be interesting to examine methods to reconstruct longer range spatial correlations that may be used to predict variable rate equations based on the current state of the tilings.  

We must ensure that all models will be consistent based on the ideas presented in section \ref{sec:constraint}.  In the current work we have not attempted to prove several natural propositions that have arisen, such as finding conditions for when a tiling scheme will be consistent.  { For example, the idea of mixed tilings raises an interesting mathematical question as to what type of {tiling}s will lead to consistent dynamics.  The triplet {scheme} that fails in the supplementary notes is a kind of mixed tiling scheme that leads to inconsistent dynamics\cite{supmat}, {\color{\red}where\sns{ }as} the mixed tiling scheme presented in section \ref{sec:RuO2} leads to consistent dynamics (tested numerically).  We conjecture that convex tilings will be necessary for consistent mixed tiling dynamics but save such investigation for future work.} { We note that even if the inconsistencies of the previous gPK models could be handled, these equations{\color{\red}\sns{s}} are typically formulated in a system of ODEs coupled to an approximate PDE that accounts for spatial correlations.  For a member of these hierarchies considering correlations of $n$ sites, there must be $n$ PDEs that must be solved (or the number of independent combinations of states considering $n$ sites; see for example \onlinecite{Mai:93,Mai:93b}); compounding this complexity is the issue of regularized, anisotropic lat{\color{\red}t}ices as we have examined in the example above on RuO$_2$ which would lead to a two dimensional PDE for each collection of state variables.  Although it is clear that in order for convergence to be achieved, less sites must be considered in a theoretically corrected von Niessen hierarchy than in the one presented in the current work, it is unclear which method would be more computationally efficient to solve due to the addition of the (potentially anisotropic) PDE.}

We have presented a generalized framework in terms of surface kinetics on square lattices.  The work immediately extends to three dimensional reaction networks and may extend to more general lattices and tiling structures.  There are many other models that take the same form of PK models such as { susceptible-infected-recovered (SIR)} models and other ecological model{s} ; indeed, pairwise models corresponding to $(2\times1)$ tilings { along with pairwise approximations for long range pairs} have been examined { in many instances \cite{KeelingEames:05,Baker:10,Baker:11,Johnston:12,Baker:13,Simpson:13}} and it will be interesting to examine whether the more generalize{d} framework presented in the current work will lead to more accurate modeling while retaining efficiency.  We remark that the current methodology may have extensions to more irregular networks { similar to the presentation found in ref. \onlinecite{Cortes:01}} and we note that this is another promising continuation of the present work.

\section{Acknowledgements}
Gregory Herschlag and Sorin Mitran GH and SM were supported by NSF-DMR 0934433.  Guang Lin would like to acknowledge support from the Applied Mathematics Program within the DOE's Office of Advanced Scientific Computing Research as part of the Collaboratory on Mathematics for Mesoscopic Modeling of Materials.  Pacific Northwest National Laboratory (PNNL) is operated by Battelle for the DOE under Contract DE-AC05-76RL01830.

%


\begin{thebibliography}{39}%
\makeatletter
\providecommand \@ifxundefined [1]{%
 \@ifx{#1\undefined}
}%
\providecommand \@ifnum [1]{%
 \ifnum #1\expandafter \@firstoftwo
 \else \expandafter \@secondoftwo
 \fi
}%
\providecommand \@ifx [1]{%
 \ifx #1\expandafter \@firstoftwo
 \else \expandafter \@secondoftwo
 \fi
}%
\providecommand \natexlab [1]{#1}%
\providecommand \enquote  [1]{``#1''}%
\providecommand \bibnamefont  [1]{#1}%
\providecommand \bibfnamefont [1]{#1}%
\providecommand \citenamefont [1]{#1}%
\providecommand \href@noop [0]{\@secondoftwo}%
\providecommand \href [0]{\begingroup \@sanitize@url \@href}%
\providecommand \@href[1]{\@@startlink{#1}\@@href}%
\providecommand \@@href[1]{\endgroup#1\@@endlink}%
\providecommand \@sanitize@url [0]{\catcode `\\12\catcode `\$12\catcode
  `\&12\catcode `\#12\catcode `\^12\catcode `\_12\catcode `\%12\relax}%
\providecommand \@@startlink[1]{}%
\providecommand \@@endlink[0]{}%
\providecommand \url  [0]{\begingroup\@sanitize@url \@url }%
\providecommand \@url [1]{\endgroup\@href {#1}{\urlprefix }}%
\providecommand \urlprefix  [0]{URL }%
\providecommand \Eprint [0]{\href }%
\providecommand \doibase [0]{http://dx.doi.org/}%
\providecommand \selectlanguage [0]{\@gobble}%
\providecommand \bibinfo  [0]{\@secondoftwo}%
\providecommand \bibfield  [0]{\@secondoftwo}%
\providecommand \translation [1]{[#1]}%
\providecommand \BibitemOpen [0]{}%
\providecommand \bibitemStop [0]{}%
\providecommand \bibitemNoStop [0]{.\EOS\space}%
\providecommand \EOS [0]{\spacefactor3000\relax}%
\providecommand \BibitemShut  [1]{\csname bibitem#1\endcsname}%
\let\auto@bib@innerbib\@empty
\bibitem [{\citenamefont {Jansen}(2012)}]{Jansen}%
  \BibitemOpen
  \bibfield  {author} {\bibinfo {author} {\bibfnamefont {A.}~\bibnamefont
  {Jansen}},\ }\href@noop {} {\emph {\bibinfo {title} {An Introduction to
  Kinetic Monte Carlo Simulations}}}\ (\bibinfo  {publisher} {Springer},\
  \bibinfo {year} {2012})\BibitemShut {NoStop}%
\bibitem [{\citenamefont {Mei}\ and\ \citenamefont {Lin}(2011)}]{MeiLin:11}%
  \BibitemOpen
  \bibfield  {author} {\bibinfo {author} {\bibfnamefont {D.}~\bibnamefont
  {Mei}}\ and\ \bibinfo {author} {\bibfnamefont {G.}~\bibnamefont {Lin}},\
  }\bibfield  {title} {\enquote {\bibinfo {title} {Effects of heat and mass
  transfer on the kinetics of co oxidation over ruo2 (110) catalyst},}\
  }\href@noop {} {\bibfield  {journal} {\bibinfo  {journal} {Catalysis Today}\
  }\textbf {\bibinfo {volume} {165}},\ \bibinfo {pages} {56--63} (\bibinfo
  {year} {2011})}\BibitemShut {NoStop}%
\bibitem [{\citenamefont {Balter}, \citenamefont {Lin},\ and\ \citenamefont
  {Tartakovsky}(2012)}]{BalterLinTartakovsky:11}%
  \BibitemOpen
  \bibfield  {author} {\bibinfo {author} {\bibfnamefont {A.}~\bibnamefont
  {Balter}}, \bibinfo {author} {\bibfnamefont {G.}~\bibnamefont {Lin}}, \ and\
  \bibinfo {author} {\bibfnamefont {A.~M.}\ \bibnamefont {Tartakovsky}},\
  }\bibfield  {title} {\enquote {\bibinfo {title} {Effect of nonlinearity in
  hybrid kinetic monte carlo-continuum models},}\ }\href@noop {} {\bibfield
  {journal} {\bibinfo  {journal} {Phys. Rev. E}\ }\textbf {\bibinfo {volume}
  {{85}}},\ \bibinfo {pages} {{016707}} (\bibinfo
  {year} {{2012}})}\BibitemShut {NoStop}%
\bibitem [{\citenamefont {Gillespie}(1992)}]{Gillespie:92}%
  \BibitemOpen
  \bibfield  {author} {\bibinfo {author} {\bibfnamefont {D.~T.}\ \bibnamefont
  {Gillespie}},\ }\bibfield  {title} {\enquote {\bibinfo {title} {A rigorous
  derivation of the chemical master equation},}\ }\href@noop {} {\bibfield
  {journal} {\bibinfo  {journal} {Physica A}\ }\textbf {\bibinfo {volume}
  {188}},\ \bibinfo {pages} {404--425} (\bibinfo {year} {1992})}\BibitemShut
  {NoStop}%
\bibitem [{\citenamefont {Karlin}(1966)}]{Karlin}%
  \BibitemOpen
  \bibfield  {author} {\bibinfo {author} {\bibfnamefont {S.}~\bibnamefont
  {Karlin}},\ }\href@noop {} {\emph {\bibinfo {title} {A First Course in
  Stochastic Processes}}}\ (\bibinfo  {publisher} {Academic Press Inc.},\
  \bibinfo {year} {1966})\BibitemShut {NoStop}%
\bibitem [{\citenamefont {Kotomin}\ and\ \citenamefont
  {Kuzovkov}(1996)}]{Kotomin:96}%
  \BibitemOpen
  \bibfield  {author} {\bibinfo {author} {\bibfnamefont {E.}~\bibnamefont
  {Kotomin}}\ and\ \bibinfo {author} {\bibfnamefont {V.}~\bibnamefont
  {Kuzovkov}},\ }\href@noop {} {\emph {\bibinfo {title} {Modern Aspects of
  Diffusion-Controlled Reactions (Comprehensive Chemical Kinetics)}}},\
  Vol.~\bibinfo {volume} {34}\ (\bibinfo  {publisher} {Elsevier},\ \bibinfo
  {address} {Amsterdam},\ \bibinfo {year} {1996})\BibitemShut {NoStop}%
\bibitem [{\citenamefont {Froment}(2005)}]{Froment:05}%
  \BibitemOpen
  \bibfield  {author} {\bibinfo {author} {\bibfnamefont {G.}~\bibnamefont
  {Froment}},\ }\bibfield  {title} {\enquote {\bibinfo {title} {Single event
  kinetic modeling of complex catalytic processes},}\ }\href@noop {} {\bibfield
   {journal} {\bibinfo  {journal} {Catal Rev Sci Eng}\ }\textbf {\bibinfo
  {volume} {47}},\ \bibinfo {pages} {83--124} (\bibinfo {year}
  {2005})}\BibitemShut {NoStop}%
\bibitem [{\citenamefont {Temel}\ \emph {et~al.}(2007)\citenamefont {Temel},
  \citenamefont {Meskine}, \citenamefont {Reuter}, \citenamefont {Scheffler},\
  and\ \citenamefont {Metiue}}]{Temeletal:07}%
  \BibitemOpen
  \bibfield  {author} {\bibinfo {author} {\bibfnamefont {B.}~\bibnamefont
  {Temel}}, \bibinfo {author} {\bibfnamefont {H.}~\bibnamefont {Meskine}},
  \bibinfo {author} {\bibfnamefont {K.}~\bibnamefont {Reuter}}, \bibinfo
  {author} {\bibfnamefont {M.}~\bibnamefont {Scheffler}}, \ and\ \bibinfo
  {author} {\bibfnamefont {H.}~\bibnamefont {Metiue}},\ }\bibfield  {title}
  {\enquote {\bibinfo {title} {Does phenomenological kinetics provide an
  adequate description of heterogeneous catalytic reactions?}}\ }\href@noop {}
  {\bibfield  {journal} {\bibinfo  {journal} {J. Chem. Phys.}\ }\textbf
  {\bibinfo {volume} {126}},\ \bibinfo {pages} {204711,1--12} (\bibinfo {year}
  {2007})}\BibitemShut {NoStop}%
\bibitem [{\citenamefont {Mai}, \citenamefont {Kuzovkov},\ and\ \citenamefont
  {{von Niessen}}(1993{\natexlab{a}})}]{Mai:93}%
  \BibitemOpen
  \bibfield  {author} {\bibinfo {author} {\bibfnamefont {J.}~\bibnamefont
  {Mai}}, \bibinfo {author} {\bibfnamefont {V.~N.}\ \bibnamefont {Kuzovkov}}, \
  and\ \bibinfo {author} {\bibfnamefont {W.}~\bibnamefont {{von Niessen}}},\
  }\bibfield  {title} {\enquote {\bibinfo {title} {A theoretical stochastic
  model for the a + $\frac{1}{2}$b$_2\rightarrow$ 0 reaction},}\ }\href@noop {}
  {\bibfield  {journal} {\bibinfo  {journal} {J. Chem. Phys.}\ }\textbf
  {\bibinfo {volume} {98}},\ \bibinfo {pages} {10017--25} (\bibinfo {year}
  {1993}{\natexlab{a}})}\BibitemShut {NoStop}%
\bibitem [{\citenamefont {Mai}, \citenamefont {Kuzovkov},\ and\ \citenamefont
  {{von Niessen}}(1993{\natexlab{b}})}]{Mai:93b}%
  \BibitemOpen
  \bibfield  {author} {\bibinfo {author} {\bibfnamefont {J.}~\bibnamefont
  {Mai}}, \bibinfo {author} {\bibfnamefont {V.~N.}\ \bibnamefont {Kuzovkov}}, \
  and\ \bibinfo {author} {\bibfnamefont {W.}~\bibnamefont {{von Niessen}}},\
  }\bibfield  {title} {\enquote {\bibinfo {title} {Stochastic model for complex
  surface-reaction systems with application to nh$_3$ formation},}\ }\href@noop
  {} {\bibfield  {journal} {\bibinfo  {journal} {Phys. Rev. E}\ }\textbf
  {\bibinfo {volume} {48}},\ \bibinfo {pages} {1700--4} (\bibinfo {year}
  {1993}{\natexlab{b}})}\BibitemShut {NoStop}%
\bibitem [{\citenamefont {Mai}, \citenamefont {Kuzovkov},\ and\ \citenamefont
  {{von Niessen}}(1994{\natexlab{a}})}]{Mai:93c}%
  \BibitemOpen
  \bibfield  {author} {\bibinfo {author} {\bibfnamefont {J.}~\bibnamefont
  {Mai}}, \bibinfo {author} {\bibfnamefont {V.~N.}\ \bibnamefont {Kuzovkov}}, \
  and\ \bibinfo {author} {\bibfnamefont {W.}~\bibnamefont {{von Niessen}}},\
  }\bibfield  {title} {\enquote {\bibinfo {title} {A general stochastic model
  for the description of the surface reaction systems},}\ }\href@noop {}
  {\bibfield  {journal} {\bibinfo  {journal} {Physica A}\ }\textbf {\bibinfo
  {volume} {203}},\ \bibinfo {pages} {298--315} (\bibinfo {year}
  {1994}{\natexlab{a}})}\BibitemShut {NoStop}%
\bibitem [{\citenamefont {Mai}, \citenamefont {Kuzovkov},\ and\ \citenamefont
  {{von Niessen}}(1994{\natexlab{b}})}]{Mai:94}%
  \BibitemOpen
  \bibfield  {author} {\bibinfo {author} {\bibfnamefont {J.}~\bibnamefont
  {Mai}}, \bibinfo {author} {\bibfnamefont {V.~N.}\ \bibnamefont {Kuzovkov}}, \
  and\ \bibinfo {author} {\bibfnamefont {W.}~\bibnamefont {{von Niessen}}},\
  }\bibfield  {title} {\enquote {\bibinfo {title} {Stochastic model for a+b 2
  surface reaction: Island formation and complete segregation},}\ }\href@noop
  {} {\bibfield  {journal} {\bibinfo  {journal} {J. Chem. Phys.}\ }\textbf
  {\bibinfo {volume} {100}},\ \bibinfo {pages} {6073--6081} (\bibinfo {year}
  {1994}{\natexlab{b}})}\BibitemShut {NoStop}%
\bibitem [{\citenamefont {Mai}, \citenamefont {Kuzovkov},\ and\ \citenamefont
  {{von Niessen}}(1994{\natexlab{c}})}]{Mai:94b}%
  \BibitemOpen
  \bibfield  {author} {\bibinfo {author} {\bibfnamefont {J.}~\bibnamefont
  {Mai}}, \bibinfo {author} {\bibfnamefont {V.~N.}\ \bibnamefont {Kuzovkov}}, \
  and\ \bibinfo {author} {\bibfnamefont {W.}~\bibnamefont {{von Niessen}}},\
  }\bibfield  {title} {\enquote {\bibinfo {title} {A simplified stochastic
  description for the a+b 2 surface reaction including a diffusion},}\
  }\href@noop {} {\bibfield  {journal} {\bibinfo  {journal} {J. Chem. Phys.}\
  }\textbf {\bibinfo {volume} {100}},\ \bibinfo {pages} {8522--8525} (\bibinfo
  {year} {1994}{\natexlab{c}})}\BibitemShut {NoStop}%
\bibitem [{\citenamefont {Mai}, \citenamefont {Kuzovkov},\ and\ \citenamefont
  {{von Niessen}}(1996)}]{Mai:96}%
  \BibitemOpen
  \bibfield  {author} {\bibinfo {author} {\bibfnamefont {J.}~\bibnamefont
  {Mai}}, \bibinfo {author} {\bibfnamefont {V.~N.}\ \bibnamefont {Kuzovkov}}, \
  and\ \bibinfo {author} {\bibfnamefont {W.}~\bibnamefont {{von Niessen}}},\
  }\bibfield  {title} {\enquote {\bibinfo {title} {A stochastic approach to
  surface reactions including energetic interactions: I. theory},}\ }\href@noop
  {} {\bibfield  {journal} {\bibinfo  {journal} {J. Phys. A. Math. Gen.}\
  }\textbf {\bibinfo {volume} {29}},\ \bibinfo {pages} {6205--6218} (\bibinfo
  {year} {1996})}\BibitemShut {NoStop}%
\bibitem [{\citenamefont {Mai}, \citenamefont {Kuzovkov},\ and\ \citenamefont
  {{von Niessen}}(1997)}]{Mai:97}%
  \BibitemOpen
  \bibfield  {author} {\bibinfo {author} {\bibfnamefont {J.}~\bibnamefont
  {Mai}}, \bibinfo {author} {\bibfnamefont {V.~N.}\ \bibnamefont {Kuzovkov}}, \
  and\ \bibinfo {author} {\bibfnamefont {W.}~\bibnamefont {{von Niessen}}},\
  }\bibfield  {title} {\enquote {\bibinfo {title} {A lotka-type model for
  oscillations in surface reactions},}\ }\href@noop {} {\bibfield  {journal}
  {\bibinfo  {journal} {J. Phys. A. Math. Gen.}\ }\textbf {\bibinfo {volume}
  {30}},\ \bibinfo {pages} {4171--4186} (\bibinfo {year} {1997})}\BibitemShut
  {NoStop}%
\bibitem [{\citenamefont {Kuzovkov}, \citenamefont {Kotomin},\ and\
  \citenamefont {{von Niessen}}(1998)}]{Kuzovkov:98}%
  \BibitemOpen
  \bibfield  {author} {\bibinfo {author} {\bibfnamefont {V.~N.}\ \bibnamefont
  {Kuzovkov}}, \bibinfo {author} {\bibfnamefont {E.~A.}\ \bibnamefont
  {Kotomin}}, \ and\ \bibinfo {author} {\bibfnamefont {W.}~\bibnamefont {{von
  Niessen}}},\ }\bibfield  {title} {\enquote {\bibinfo {title}
  {Discrete-lattice theory for frankel-defect aggregation in irradiated ionic
  solids},}\ }\href@noop {} {\bibfield  {journal} {\bibinfo  {journal} {Phys.
  Rev. B}\ }\textbf {\bibinfo {volume} {58}},\ \bibinfo {pages} {8454--8463}
  (\bibinfo {year} {1998})}\BibitemShut {NoStop}%
\bibitem [{\citenamefont {Dickman}\ \emph {et~al.}(1999)\citenamefont
  {Dickman}, \citenamefont {Grandi}, \citenamefont {Figueiredo},\ and\
  \citenamefont {Dickman}}]{Dickman:99}%
  \BibitemOpen
  \bibfield  {author} {\bibinfo {author} {\bibfnamefont {A.~G.}\ \bibnamefont
  {Dickman}}, \bibinfo {author} {\bibfnamefont {B.~C.~S.}\ \bibnamefont
  {Grandi}}, \bibinfo {author} {\bibfnamefont {W.}~\bibnamefont {Figueiredo}},
  \ and\ \bibinfo {author} {\bibfnamefont {R.}~\bibnamefont {Dickman}},\
  }\bibfield  {title} {\enquote {\bibinfo {title} {Theory of the no+co
  surface-reaction model},}\ }\href@noop {} {\bibfield  {journal} {\bibinfo
  {journal} {Phys. Rev. E}\ }\textbf {\bibinfo {volume} {59}},\ \bibinfo
  {pages} {6361--6369} (\bibinfo {year} {1999})}\BibitemShut {NoStop}%
\bibitem [{\citenamefont {Cort\'{e}s}\ \emph {et~al.}(2003)\citenamefont
  {Cort\'{e}s}, \citenamefont {Narv\'{a}ez}, \citenamefont {Puschmann},\ and\
  \citenamefont {Valencia}}]{Cortes:01}%
  \BibitemOpen
  \bibfield  {author} {\bibinfo {author} {\bibfnamefont {J.}~\bibnamefont
  {Cort\'{e}s}}, \bibinfo {author} {\bibfnamefont {A.}~\bibnamefont
  {Narv\'{a}ez}}, \bibinfo {author} {\bibfnamefont {H.}~\bibnamefont
  {Puschmann}}, \ and\ \bibinfo {author} {\bibfnamefont {E.}~\bibnamefont
  {Valencia}},\ }\bibfield  {title} {\enquote {\bibinfo {title} {Mean field
  theory studies of surface reactions on disordered substrates},}\ }\href@noop
  {} {\bibfield  {journal} {\bibinfo  {journal} {Chem. Phys.}\ }\textbf
  {\bibinfo {volume} {288}},\ \bibinfo {pages} {77--88} (\bibinfo {year}
  {2003})}\BibitemShut {NoStop}%
\bibitem [{\citenamefont {{De Decker}}\ \emph {et~al.}(2002)\citenamefont {{De
  Decker}}, \citenamefont {Baras}, \citenamefont {Kruse},\ and\ \citenamefont
  {Nicolis}}]{DeDecker:02}%
  \BibitemOpen
  \bibfield  {author} {\bibinfo {author} {\bibfnamefont {Y.}~\bibnamefont {{De
  Decker}}}, \bibinfo {author} {\bibfnamefont {F.}~\bibnamefont {Baras}},
  \bibinfo {author} {\bibfnamefont {N.}~\bibnamefont {Kruse}}, \ and\ \bibinfo
  {author} {\bibfnamefont {G.}~\bibnamefont {Nicolis}},\ }\bibfield  {title}
  {\enquote {\bibinfo {title} {Modeling the no+h 2 reaction on a pt field
  emitter tip: Mean-field analysis and monte carlo simulations},}\ }\href@noop
  {} {\bibfield  {journal} {\bibinfo  {journal} {J. Chem. Phys.}\ }\textbf
  {\bibinfo {volume} {117}},\ \bibinfo {pages} {10244--10257} (\bibinfo {year}
  {2002})}\BibitemShut {NoStop}%
\bibitem [{\citenamefont {Baker}\ and\ \citenamefont
  {Simpson}(2010)}]{Baker:10}%
  \BibitemOpen
  \bibfield  {author} {\bibinfo {author} {\bibfnamefont {R.~E.}\ \bibnamefont
  {Baker}}\ and\ \bibinfo {author} {\bibfnamefont {M.~J.}\ \bibnamefont
  {Simpson}},\ }\bibfield  {title} {\enquote {\bibinfo {title} {Correcting
  mean-field approximations for birth-death-movement processes},}\ }\href@noop
  {} {\bibfield  {journal} {\bibinfo  {journal} {Phys. Rev. E}\ }\textbf
  {\bibinfo {volume} {82}},\ \bibinfo {pages} {041905} (\bibinfo {year}
  {2010})}\BibitemShut {NoStop}%
\bibitem [{\citenamefont {Simpson}\ and\ \citenamefont
  {Baker}(2011)}]{Baker:11}%
  \BibitemOpen
  \bibfield  {author} {\bibinfo {author} {\bibfnamefont {M.~J.}\ \bibnamefont
  {Simpson}}\ and\ \bibinfo {author} {\bibfnamefont {R.~E.}\ \bibnamefont
  {Baker}},\ }\bibfield  {title} {\enquote {\bibinfo {title} {Corrected
  mean-field models for spatially dependent advection-diffusion-reaction
  phenomena},}\ }\href@noop {} {\bibfield  {journal} {\bibinfo  {journal}
  {Phys. Rev. E}\ }\textbf {\bibinfo {volume} {83}},\ \bibinfo {pages} {051922}
  (\bibinfo {year} {2011})}\BibitemShut {NoStop}%
\bibitem [{\citenamefont {Johnston}, \citenamefont {Simpson},\ and\
  \citenamefont {Baker}(2012)}]{Johnston:12}%
  \BibitemOpen
  \bibfield  {author} {\bibinfo {author} {\bibfnamefont {S.}~\bibnamefont
  {Johnston}}, \bibinfo {author} {\bibfnamefont {M.~J.}\ \bibnamefont
  {Simpson}}, \ and\ \bibinfo {author} {\bibfnamefont {R.~E.}\ \bibnamefont
  {Baker}},\ }\bibfield  {title} {\enquote {\bibinfo {title} {Mean-field
  descriptions of collective migration with strong adhesion},}\ }\href@noop {}
  {\bibfield  {journal} {\bibinfo  {journal} {Chem. Engin. Sc.}\ }\textbf
  {\bibinfo {volume} {85}},\ \bibinfo {pages} {051922} (\bibinfo {year}
  {2012})}\BibitemShut {NoStop}%
\bibitem [{\citenamefont {Markham}, \citenamefont {Simpson},\ and\
  \citenamefont {Baker}(2013)}]{Baker:13}%
  \BibitemOpen
  \bibfield  {author} {\bibinfo {author} {\bibfnamefont {D.~C.}\ \bibnamefont
  {Markham}}, \bibinfo {author} {\bibfnamefont {M.~J.}\ \bibnamefont
  {Simpson}}, \ and\ \bibinfo {author} {\bibfnamefont {R.~E.}\ \bibnamefont
  {Baker}},\ }\bibfield  {title} {\enquote {\bibinfo {title} {Simplified method
  for including spatial correlations in mean-field approximations},}\
  }\href@noop {} {\bibfield  {journal} {\bibinfo  {journal} {Phys. Rev. E}\
  }\textbf {\bibinfo {volume} {87}},\ \bibinfo {pages} {062702} (\bibinfo
  {year} {2013})}\BibitemShut {NoStop}%
\bibitem [{\citenamefont {Simpson}\ \emph {et~al.}(2013)\citenamefont
  {Simpson}, \citenamefont {Binder}, \citenamefont {Haridas}, \citenamefont
  {Wood}, \citenamefont {Treloar}, \citenamefont {McElwain},\ and\
  \citenamefont {Baker}}]{Simpson:13}%
  \BibitemOpen
  \bibfield  {author} {\bibinfo {author} {\bibfnamefont {M.}~\bibnamefont
  {Simpson}}, \bibinfo {author} {\bibfnamefont {B.~J.}\ \bibnamefont {Binder}},
  \bibinfo {author} {\bibfnamefont {P.}~\bibnamefont {Haridas}}, \bibinfo
  {author} {\bibfnamefont {B.~K.}\ \bibnamefont {Wood}}, \bibinfo {author}
  {\bibfnamefont {K.~K.}\ \bibnamefont {Treloar}}, \bibinfo {author}
  {\bibfnamefont {D.~L.~S.}\ \bibnamefont {McElwain}}, \ and\ \bibinfo {author}
  {\bibfnamefont {R.~E.}\ \bibnamefont {Baker}},\ }\bibfield  {title} {\enquote
  {\bibinfo {title} {Experimental and modelling investigation of monolayer
  development with clustering},}\ }\href@noop {} {\bibfield  {journal}
  {\bibinfo  {journal} {Bull. of Math. Biol.}\ }\textbf {\bibinfo {volume}
  {75}},\ \bibinfo {pages} {871--889} (\bibinfo {year} {2013})}\BibitemShut
  {NoStop}%
\bibitem [{\citenamefont {Hansen}\ and\ \citenamefont
  {Neurock}(1999)}]{Hansen:99}%
  \BibitemOpen
  \bibfield  {author} {\bibinfo {author} {\bibfnamefont {E.}~\bibnamefont
  {Hansen}}\ and\ \bibinfo {author} {\bibfnamefont {M.}~\bibnamefont
  {Neurock}},\ }\bibfield  {title} {\enquote {\bibinfo {title} {Modeling
  surface kinetics with first-principles-based molecular simulation},}\
  }\href@noop {} {\bibfield  {journal} {\bibinfo  {journal} {Chem. Engin. Sc.}\
  }\textbf {\bibinfo {volume} {54}},\ \bibinfo {pages} {3411--3421} (\bibinfo
  {year} {1999})}\BibitemShut {NoStop}%
\bibitem [{\citenamefont {Zeigarnik}\ \emph {et~al.}(2003)\citenamefont
  {Zeigarnik}, \citenamefont {Abramova}, \citenamefont {Baronov},\ and\
  \citenamefont {Shustorovich}}]{Zeigarnik:03}%
  \BibitemOpen
  \bibfield  {author} {\bibinfo {author} {\bibfnamefont {A.~V.}\ \bibnamefont
  {Zeigarnik}}, \bibinfo {author} {\bibfnamefont {L.~A.}\ \bibnamefont
  {Abramova}}, \bibinfo {author} {\bibfnamefont {S.}~\bibnamefont {Baronov}}, \
  and\ \bibinfo {author} {\bibfnamefont {E.}~\bibnamefont {Shustorovich}},\
  }\bibfield  {title} {\enquote {\bibinfo {title} {Monte carlo modeling of
  ubi-qep coverage-dependent atomic chemisorption},}\ }\href@noop {} {\bibfield
   {journal} {\bibinfo  {journal} {Surface Science}\ }\textbf {\bibinfo
  {volume} {541}},\ \bibinfo {pages} {76--90} (\bibinfo {year}
  {2003})}\BibitemShut {NoStop}%
\bibitem [{\citenamefont {Keiken}, \citenamefont {Neurock},\ and\ \citenamefont
  {Mei}(2005)}]{Kieken:04}%
  \BibitemOpen
  \bibfield  {author} {\bibinfo {author} {\bibfnamefont {L.}~\bibnamefont
  {Keiken}}, \bibinfo {author} {\bibfnamefont {M.}~\bibnamefont {Neurock}}, \
  and\ \bibinfo {author} {\bibfnamefont {D.}~\bibnamefont {Mei}},\ }\bibfield
  {title} {\enquote {\bibinfo {title} {Screening by kinetic monte carlo
  simulation of pt-au(100) surfaces for the steady-state decomposition of
  nitric oxide in excess dioxygen},}\ }\href@noop {} {\bibfield  {journal}
  {\bibinfo  {journal} {J. Phys. Chem. B}\ }\textbf {\bibinfo {volume} {109}},\
  \bibinfo {pages} {2234--2244} (\bibinfo {year} {2005})}\BibitemShut {NoStop}%
\bibitem [{\citenamefont {Mei}\ \emph {et~al.}(2004)\citenamefont {Mei},
  \citenamefont {Ge}, \citenamefont {Neurock}, \citenamefont {Kieken},\ and\
  \citenamefont {Lerou}}]{Mei:06}%
  \BibitemOpen
  \bibfield  {author} {\bibinfo {author} {\bibfnamefont {D.}~\bibnamefont
  {Mei}}, \bibinfo {author} {\bibfnamefont {Q.}~\bibnamefont {Ge}}, \bibinfo
  {author} {\bibfnamefont {M.}~\bibnamefont {Neurock}}, \bibinfo {author}
  {\bibfnamefont {L.}~\bibnamefont {Kieken}}, \ and\ \bibinfo {author}
  {\bibfnamefont {J.}~\bibnamefont {Lerou}},\ }\bibfield  {title} {\enquote
  {\bibinfo {title} {First-principles-based kinetic monte carlo simulation of
  nitric oxide decomposition over pt and rh surfaces under lean-burn
  conditions},}\ }\href@noop {} {\bibfield  {journal} {\bibinfo  {journal}
  {Mol. Phys.}\ ,\ \bibinfo {pages} {361--369}} (\bibinfo {year}
  {2004})}\BibitemShut {NoStop}%
\bibitem [{\citenamefont {Reuter}\ and\ \citenamefont
  {Scheffler}(2006)}]{ReuterScheffler:06}%
  \BibitemOpen
  \bibfield  {author} {\bibinfo {author} {\bibfnamefont {K.}~\bibnamefont
  {Reuter}}\ and\ \bibinfo {author} {\bibfnamefont {M.}~\bibnamefont
  {Scheffler}},\ }\bibfield  {title} {\enquote {\bibinfo {title}
  {First-principles kinetic monte carlo simulations for heterogeneous
  catalysis: Application to the $co$ oxidation at $ruo_2(110)$},}\ }\href@noop
  {} {\bibfield  {journal} {\bibinfo  {journal} {Phys. Rev. B}\ }\textbf
  {\bibinfo {volume} {73}},\ \bibinfo {pages} {045433,1--17} (\bibinfo {year}
  {2006})}\BibitemShut {NoStop}%
\bibitem [{\citenamefont {Rogal}, \citenamefont {Reuter},\ and\ \citenamefont
  {Scheffler}(2007)}]{Rogal:07}%
  \BibitemOpen
  \bibfield  {author} {\bibinfo {author} {\bibfnamefont {J.}~\bibnamefont
  {Rogal}}, \bibinfo {author} {\bibfnamefont {K.}~\bibnamefont {Reuter}}, \
  and\ \bibinfo {author} {\bibfnamefont {M.}~\bibnamefont {Scheffler}},\
  }\bibfield  {title} {\enquote {\bibinfo {title} {First-principles statistical
  mechanics study of the stability of a subnanometer thin surface ?oxide in
  reactive environments: Co oxidation at pd(100)},}\ }\href@noop {} {\bibfield
  {journal} {\bibinfo  {journal} {Phys. Rev. Lett.}\ }\textbf {\bibinfo
  {volume} {98}},\ \bibinfo {pages} {046101} (\bibinfo {year}
  {2007})}\BibitemShut {NoStop}%
\bibitem [{\citenamefont {Herdrich}\ \emph {et~al.}(2012)\citenamefont
  {Herdrich}, \citenamefont {Fertig}, \citenamefont {Petkow}, \citenamefont
  {Steinbeck},\ and\ \citenamefont {Fasoulas}}]{Herdrich:12}%
  \BibitemOpen
  \bibfield  {author} {\bibinfo {author} {\bibfnamefont {G.}~\bibnamefont
  {Herdrich}}, \bibinfo {author} {\bibfnamefont {M.}~\bibnamefont {Fertig}},
  \bibinfo {author} {\bibfnamefont {D.}~\bibnamefont {Petkow}}, \bibinfo
  {author} {\bibfnamefont {A.}~\bibnamefont {Steinbeck}}, \ and\ \bibinfo
  {author} {\bibfnamefont {S.}~\bibnamefont {Fasoulas}},\ }\bibfield  {title}
  {\enquote {\bibinfo {title} {Experimental and numerical techniques to assess
  catalysis},}\ }\href@noop {} {\bibfield  {journal} {\bibinfo  {journal}
  {Topics in Catalysis}\ }\textbf {\bibinfo {volume} {48-49}},\ \bibinfo
  {pages} {27--41} (\bibinfo {year} {2012})}\BibitemShut {NoStop}%
\bibitem [{\citenamefont {Liu}\ and\ \citenamefont {Evans}(2013)}]{Liu:13}%
  \BibitemOpen
  \bibfield  {author} {\bibinfo {author} {\bibfnamefont {D.-J.}\ \bibnamefont
  {Liu}}\ and\ \bibinfo {author} {\bibfnamefont {J.~W.}\ \bibnamefont
  {Evans}},\ }\bibfield  {title} {\enquote {\bibinfo {title} {Realistic
  multisite lattice-gas modeling and kmc simulation of catalytic surface
  reactions: Kinetics and multiscale spatial behavior for co-oxidation on metal
  (100) surfaces},}\ }\href@noop {} {\bibfield  {journal} {\bibinfo  {journal}
  {Progress in Surface Science}\ }\textbf {\bibinfo {volume} {88}},\ \bibinfo
  {pages} {393--521} (\bibinfo {year} {2013})}\BibitemShut {NoStop}%
\bibitem [{\citenamefont {Hoffmann}\ and\ \citenamefont
  {Reuter}(2013)}]{Hoffmann:13}%
  \BibitemOpen
  \bibfield  {author} {\bibinfo {author} {\bibfnamefont {M.~J.}\ \bibnamefont
  {Hoffmann}}\ and\ \bibinfo {author} {\bibfnamefont {K.}~\bibnamefont
  {Reuter}},\ }\bibfield  {title} {\enquote {\bibinfo {title} {Co oxidation on
  pd(100) versus pdo(101)- $(\sqrt{5}\times\sqrt{5})r27^o$: First-principles
  kinetic phase diagrams and bistability conditions},}\ }\href@noop {}
  {\bibfield  {journal} {\bibinfo  {journal} {Topics in Catalysis}\ }\textbf
  {\bibinfo {volume} {57}},\ \bibinfo {pages} {159--170} (\bibinfo {year}
  {2013})}\BibitemShut {NoStop}%
\bibitem [{\citenamefont {Callejas-Tovara}\ \emph {et~al.}(2013)\citenamefont
  {Callejas-Tovara}, \citenamefont {Diaza}, \citenamefont {de~la Hoza},\ and\
  \citenamefont {Balbuena}}]{Callejas:13}%
  \BibitemOpen
  \bibfield  {author} {\bibinfo {author} {\bibfnamefont {R.}~\bibnamefont
  {Callejas-Tovara}}, \bibinfo {author} {\bibfnamefont {C.~A.}\ \bibnamefont
  {Diaza}}, \bibinfo {author} {\bibfnamefont {J.~M.~M.}\ \bibnamefont {de~la
  Hoza}}, \ and\ \bibinfo {author} {\bibfnamefont {P.~B.}\ \bibnamefont
  {Balbuena}},\ }\bibfield  {title} {\enquote {\bibinfo {title} {Dealloying of
  platinum-based alloy catalysts: Kinetic monte carlo simulations},}\
  }\href@noop {} {\bibfield  {journal} {\bibinfo  {journal} {Electrochimica
  Acta}\ }\textbf {\bibinfo {volume} {101}},\ \bibinfo {pages} {326--333}
  (\bibinfo {year} {2013})}\BibitemShut {NoStop}%
\bibitem [{\citenamefont {Gibson}, \citenamefont {Killeen},\ and\ \citenamefont
  {Sibener}(2014)}]{Gibson:14}%
  \BibitemOpen
  \bibfield  {author} {\bibinfo {author} {\bibfnamefont {K.~D.}\ \bibnamefont
  {Gibson}}, \bibinfo {author} {\bibfnamefont {D.~R.}\ \bibnamefont {Killeen}},
  \ and\ \bibinfo {author} {\bibfnamefont {S.}~\bibnamefont {Sibener}},\
  }\bibfield  {title} {\enquote {\bibinfo {title} {Comparison of the surface
  and subsurface oxygen reactivity and dynamics with co adsorbed on rh(111)},}\
  }\href@noop {} {\bibfield  {journal} {\bibinfo  {journal} {J. Phys. Chem. C}\
  }\textbf {\bibinfo {volume} {118}},\ \bibinfo {pages} {14977--14982}
  (\bibinfo {year} {2014})}\BibitemShut {NoStop}%
\bibitem [{sup()}]{supmat}%
  \BibitemOpen
  \href@noop {} {\bibinfo  {journal} {See Supplementary Material Document
  No.\_\_\_\_\_\_\_\_\_ for an example of an inconsistent triplet model. For
  information on Supplementary Material, see
  http://www.aip.org/pubservs/epaps.html}\ }\BibitemShut {NoStop}%
\bibitem [{\citenamefont {Jones}\ \emph {et~al.}(01  )\citenamefont {Jones},
  \citenamefont {Oliphant}, \citenamefont {Peterson} \emph {et~al.}}]{Scipy}%
  \BibitemOpen
\bibfield  {journal} {  }\bibfield  {author} {\bibinfo {author} {\bibfnamefont
  {E.}~\bibnamefont {Jones}}, \bibinfo {author} {\bibfnamefont
  {T.}~\bibnamefont {Oliphant}}, \bibinfo {author} {\bibfnamefont
  {P.}~\bibnamefont {Peterson}},  \emph {et~al.},\ }\href
  {http://www.scipy.org/} {\enquote {\bibinfo {title} {{SciPy}: Open source
  scientific tools for {Python}},}\ } (\bibinfo {year} {2001--})\BibitemShut
  {NoStop}%
\bibitem [{\citenamefont {Matera}, \citenamefont {Meskine},\ and\ \citenamefont
  {Reuter}(2011)}]{Matera:11}%
  \BibitemOpen
  \bibfield  {author} {\bibinfo {author} {\bibfnamefont {S.}~\bibnamefont
  {Matera}}, \bibinfo {author} {\bibfnamefont {H.}~\bibnamefont {Meskine}}, \
  and\ \bibinfo {author} {\bibfnamefont {K.}~\bibnamefont {Reuter}},\
  }\bibfield  {title} {\enquote {\bibinfo {title} {Adlayer inhomogeneity
  without lateral interactions: Rationalizing correlation effects in co
  oxidation at ruo2(110) with first-principles kinetic monte carlo},}\
  }\href@noop {} {\bibfield  {journal} {\bibinfo  {journal} {J. Chem. Phys.}\
  }\textbf {\bibinfo {volume} {134}},\ \bibinfo {pages} {064713} (\bibinfo
  {year} {2011})}\BibitemShut {NoStop}%
\bibitem [{\citenamefont {Keeling}\ and\ \citenamefont
  {Eames}(2005)}]{KeelingEames:05}%
  \BibitemOpen
  \bibfield  {author} {\bibinfo {author} {\bibfnamefont {M.}~\bibnamefont
  {Keeling}}\ and\ \bibinfo {author} {\bibfnamefont {K.}~\bibnamefont
  {Eames}},\ }\bibfield  {title} {\enquote {\bibinfo {title} {Networks and
  epidemic models},}\ }\href@noop {} {\bibfield  {journal} {\bibinfo  {journal}
  {J. Roy. Soc.}\ }\textbf {\bibinfo {volume} {2}},\ \bibinfo {pages}
  {295--307} (\bibinfo {year} {2005})}\BibitemShut {NoStop}%
\end{thebibliography}

\end{document}